\documentclass[article, nojss]{jss}
\usepackage[utf8]{inputenc}

\author{
John Ehrlinger\\Microsoft
}
\title{\pkg{ggRandomForests}: Exploring Random Forest Survival}
\Keywords{random forest, survival, vimp, minimal depth, \proglang{R}, \pkg{randomForestSRC}, \pkg{ggRandomForests}, \pkg{randomForest}}

\Abstract{
Random forest \citep{Breiman:2001} (RF) is a non-parametric statistical
method requiring no distributional assumptions on covariate relation to
the response. RF is a robust, nonlinear technique that optimizes
predictive accuracy by fitting an ensemble of trees to stabilize model
estimates. Random survival forests (RSF)
\citep{Ishwaran:2007a, Ishwaran:2008} are an extension of Breimans RF
techniques allowing efficient non-parametric analysis of time to event
data. The \pkg{randomForestSRC} package \citep{Ishwaran:RFSRC:2014} is a
unified treatment of Breimans random forest for survival, regression and
classification problems.

Predictive accuracy makes RF an attractive alternative to parametric
models, though complexity and interpretability of the forest hinder
wider application of the method. We introduce the \pkg{ggRandomForests}
package, tools for visually understand random forest models grown in
\proglang{R} \citep{rcore} with the \pkg{randomForestSRC} package. The
\pkg{ggRandomForests} package is structured to extract intermediate data
objects from \pkg{randomForestSRC} objects and generate figures using
the \pkg{ggplot2} \citep{Wickham:2009} graphics package.

This document is structured as a tutorial for building random forest for
survival with the \pkg{randomForestSRC} package and using the
\pkg{ggRandomForests} package for investigating how the forest is
constructed. We analyse the Primary Biliary Cirrhosis of the liver data
from a clinical trial at the Mayo Clinic \citep{fleming:1991}. We
demonstrate random forest variable selection using Variable Importance
(VIMP) \citep{Breiman:2001} and Minimal Depth \citep{Ishwaran:2010}, a
property derived from the construction of each tree within the forest.
We will also demonstrate the use of variable dependence and partial
dependence plots \citep{Friedman:2000} to aid in the interpretation of
RSF results. We then examine variable interactions between covariates
using conditional variable dependence plots. Our aim is to demonstrate
the strength of using Random Forest methods for both prediction and
information retrieval, specifically in time to event data settings.
}

\Plainauthor{John Ehrlinger}
\Plaintitle{ggRandomForests: Exploring Random Forest Survival}
\Shorttitle{\pkg{ggRandomForests}}
\Plainkeywords{random forest, survival, vimp, minimal depth, r, randomForestSRC, ggRandomForests, randomForest}

\Submitdate{}

\Address{
    John Ehrlinger\\
  Microsoft\\
  One Memorial Drive Cambridge, MA 02142\\
  E-mail: \href{mailto:john.ehrlinger@gmail.com}{\nolinkurl{john.ehrlinger@gmail.com}}\\
  URL: \url{https://github.com/ehrlinger/ggRandomForests}\\~\\
  }

\usepackage{amssymb, amsmath, booktabs, longtable}

\begin{document}

\section{Introduction}\label{introduction}

Random forest \citep{Breiman:2001} (RF) is a non-parametric statistical
method which requires no distributional assumptions on covariate
relation to the response. RF is a robust, nonlinear technique that
optimizes predictive accuracy by fitting an ensemble of trees to
stabilize model estimates. Random Survival Forest (RSF)
\citep{Ishwaran:2007a, Ishwaran:2008} is an extension of Breiman's RF
techniques to survival settings, allowing efficient non-parametric
analysis of time to event data. The \pkg{randomForestSRC} package
(\url{http://CRAN.R-project.org/package=randomForestSRC})
\citep{Ishwaran:RFSRC:2014} is a unified treatment of Breiman's random
forest for survival, regression and classification problems.

Predictive accuracy make RF an attractive alternative to parametric
models, though complexity and interpretability of the forest hinder
wider application of the method. We introduce the \pkg{ggRandomForests}
package (\url{http://CRAN.R-project.org/package=ggRandomForests}) for
visually exploring random forest models. The \pkg{ggRandomForests}
package is structured to extract intermediate data objects from
\pkg{randomForestSRC} objects and generate figures using the
\pkg{ggplot2} graphics package
(\url{http://CRAN.R-project.org/package=ggplot2}) \citep{Wickham:2009}.

Many of the figures created by the \pkg{ggRandomForests} package are
also available directly from within the \pkg{randomForestSRC} package.
However \pkg{ggRandomForests} offers the following advantages:

\begin{itemize}
\item
  Separation of data and figures: \pkg{ggRandomForests} contains
  functions that operate on either the \texttt{rfsrc} forest object
  directly, or on the output from \pkg{randomForestSRC} post processing
  functions (i.e., \texttt{plot.variable}, \texttt{var.select}) to
  generate intermediate \pkg{ggRandomForests} data objects.
  \pkg{ggRandomForests} functions are provide to further process these
  objects and plot results using the \pkg{ggplot2} graphics package.
  Alternatively, users can use these data objects for their own custom
  plotting or analysis operations.
\item
  Each data object/figure is a single, self contained unit. This allows
  simple modification and manipulation of the data or \texttt{ggplot}
  objects to meet users specific needs and requirements.
\item
  We chose to use the \pkg{ggplot2} package for our figures for
  flexibility in modifying the output. Each \pkg{ggRandomForests} plot
  function returns either a single \texttt{ggplot} object, or a
  \texttt{list} of \texttt{ggplot} objects, allowing the use of
  additional \pkg{ggplot2} functions to modify and customize the final
  figures.
\end{itemize}

This document is structured as a tutorial for using the
\pkg{randomForestSRC} package for building and post-processing random
survival forest models and using the \pkg{ggRandomForests} package for
understanding how the forest is constructed. In this tutorial, we will
build a random survival forest for the primary biliary cirrhosis (PBC)
of the liver data set \citep{fleming:1991}, available in the
\pkg{randomForestSRC} package.

In \autoref{data-summary-primary-biliary-cirrhosis-pbc-data-set} we
introduce the \texttt{pbc} data set and summarize the proportional
hazards analysis of this data from Chapter 4 of \citep{fleming:1991}. In
\autoref{random-survival-forest}, we describe how to grow a random
survival forest with the \pkg{randomForestSRC} package. Random forest is
not a parsimonious method, but uses all variables available in the data
set to construct the response predictor. We demonstrate random forest
variable selection techniques (\autoref{variable-selection}) using
Variable Importance (VIMP) \citep{Breiman:2001} in
\autoref{variable-importance} and Minimal Depth \citep{Ishwaran:2010} in
\autoref{minimal-depth}. We then compare both methods with variables
used in the \citep{fleming:1991} model.

Once we have an idea of which variables we are most interested in, we
use dependence plots \citep{Friedman:2000}
(\autoref{variableresponse-dependence}) to understand how these
variables are related to the response. Variable dependence
(\autoref{variable-dependence}) plots give us an idea of the overall
trend of a variable/response relation, while partial dependence plots
(\autoref{partial-dependence}) show us the risk adjusted relation by
averaging out the effects of other variables. Dependence plots often
show strongly non-linear variable/response relations that are not easily
obtained through parametric modeling.

We then graphically examine forest variable interactions with the use of
variable and partial dependence conditioning plots (coplots)
\citep{chambers:1992, cleveland:1993}
(\autoref{conditional-dependence-plots}) and the analogouse partial
dependence surfaces (\autoref{partial-plot-surfaces}) before adding
concluding remarks in \autoref{conclusion}.

\section{Data summary: primary biliary cirrhosis (PBC) data
set}\label{data-summary-primary-biliary-cirrhosis-pbc-data-set}

The \emph{primary biliary cirrhosis} of the liver (PBC) study consists
of 424 PBC patients referred to Mayo Clinic between 1974 and 1984 who
met eligibility criteria for a randomized placebo controlled trial of
the drug D-penicillamine (DPCA). The data is described in
\cite[Chapter 0.2]{fleming:1991} and a partial likelihood model (Cox
proportional hazards) is developed in Chapter 4.4. The \texttt{pbc} data
set, included in the \pkg{randomForestSRC} package, contains 418
observations, of which 312 patients participated in the randomized trial
\cite[Appendix D]{fleming:1991}.

\begin{Schunk}
\begin{Sinput}
R> data("pbc", package = "randomForestSRC")
\end{Sinput}
\end{Schunk}

For this analysis, we modify some of the data for better formatting of
our results. Since the data contains about 12 years of follow up, we
prefer using \texttt{years} instead of \texttt{days} to describe
survival. We also convert the \texttt{age} variable to years, and the
\texttt{treatment} variable to a factor containing levels of
\texttt{c("DPCA",\ "placebo")}. The variable names, type and description
are given in \autoref{T:dataLabs}.

\begin{Schunk}
\begin{table}

\caption{\label{tab:dta-table}`pbc` data set variable dictionary.\label{T:dataLabs}}
\centering
\begin{tabular}[t]{l|l|l}
\hline
Variable name & Description & Type\\
\hline
years & Time (years) & numeric\\
\hline
status & Event (F = censor, T = death) & logical\\
\hline
treatment & Treament (DPCA, Placebo) & factor\\
\hline
age & Age (years) & numeric\\
\hline
sex & Female = T & logical\\
\hline
ascites & Presence of Asictes & logical\\
\hline
hepatom & Presence of Hepatomegaly & logical\\
\hline
spiders & Presence of Spiders & logical\\
\hline
edema & Edema (0, 0.5, 1) & factor\\
\hline
bili & Serum Bilirubin (mg/dl) & numeric\\
\hline
chol & Serum Cholesterol (mg/dl) & integer\\
\hline
albumin & Albumin (gm/dl) & numeric\\
\hline
copper & Urine Copper (ug/day) & integer\\
\hline
alk & Alkaline Phosphatase (U/liter) & numeric\\
\hline
sgot & SGOT (U/ml) & numeric\\
\hline
trig & Triglicerides (mg/dl) & integer\\
\hline
platelet & Platelets per cubic ml/1000 & integer\\
\hline
prothrombin & Prothrombin time (sec) & numeric\\
\hline
stage & Histologic Stage & factor\\
\hline
\end{tabular}
\end{table}

\end{Schunk}

\subsection{Exploratory data analysis}\label{exploratory-data-analysis}

It is good practice to view your data before beginning analysis.
Exploratory Data Analysis (EDA) \citep{Tukey:1977} will help you to
understand the data, and find outliers, missing values and other data
anomalies within each variable before getting deep into the analysis. To
this end, we use \pkg{ggplot2} figures with the \texttt{facet\_wrap}
function to create two sets of panel plots, one of histograms for
categorical variables (\autoref{fig:categoricalEDA}), and another of
scatter plots for continuous variables (\autoref{fig:continuousEDA}).
Variables are plotted along a continuous variable on the X-axis to
separate the individual observations.

\begin{Schunk}
\begin{figure}[!htb]

{\centering \includegraphics{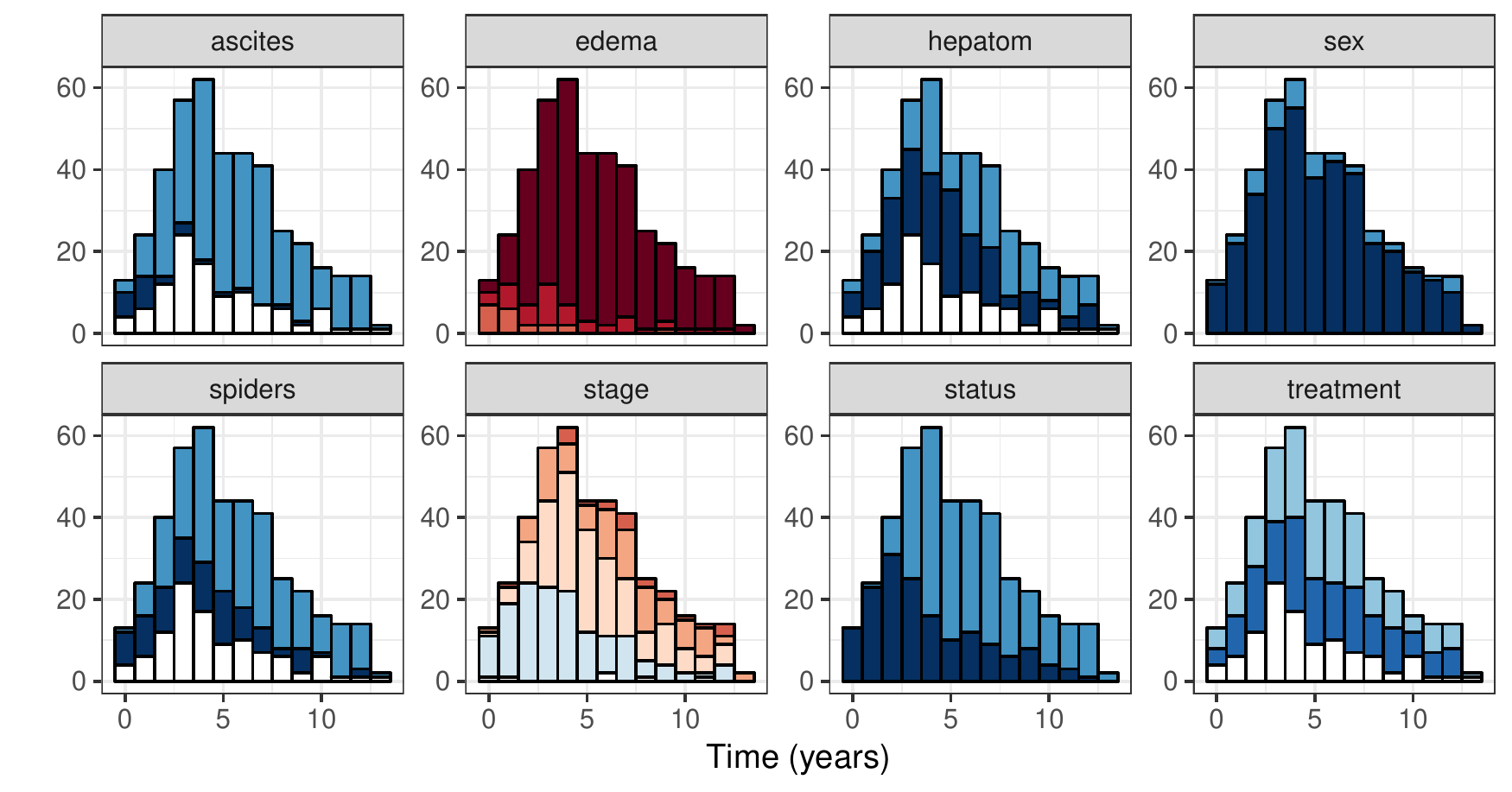}

}

\caption[EDA plots for categorical variables (logicals and factors)]{EDA plots for categorical variables (logicals and factors). Bars indicate number of patients within 1 year of followup interval for each categorical variable. Colors correspond to class membership within each variable. Missing values are included in white.}\label{fig:categoricalEDA}
\end{figure}
\end{Schunk}

In categorical EDA plots (\autoref{fig:categoricalEDA}), we are looking
for patterns of missing data (white portion of bars). We often use
surgical date for our X-axis variable to look for possible periods of
low enrollment. There is not a comparable variable available in the
\texttt{pbc} data set, so instead we used follow up time
(\texttt{years}). Another reasonable choice may have been to use the
patient \texttt{age} variable for the X-axis. The important quality of
the selected variable is to spread the observations out to aid in
finding data anomalies.

\begin{Schunk}
\begin{figure}[!htb]

{\centering \includegraphics{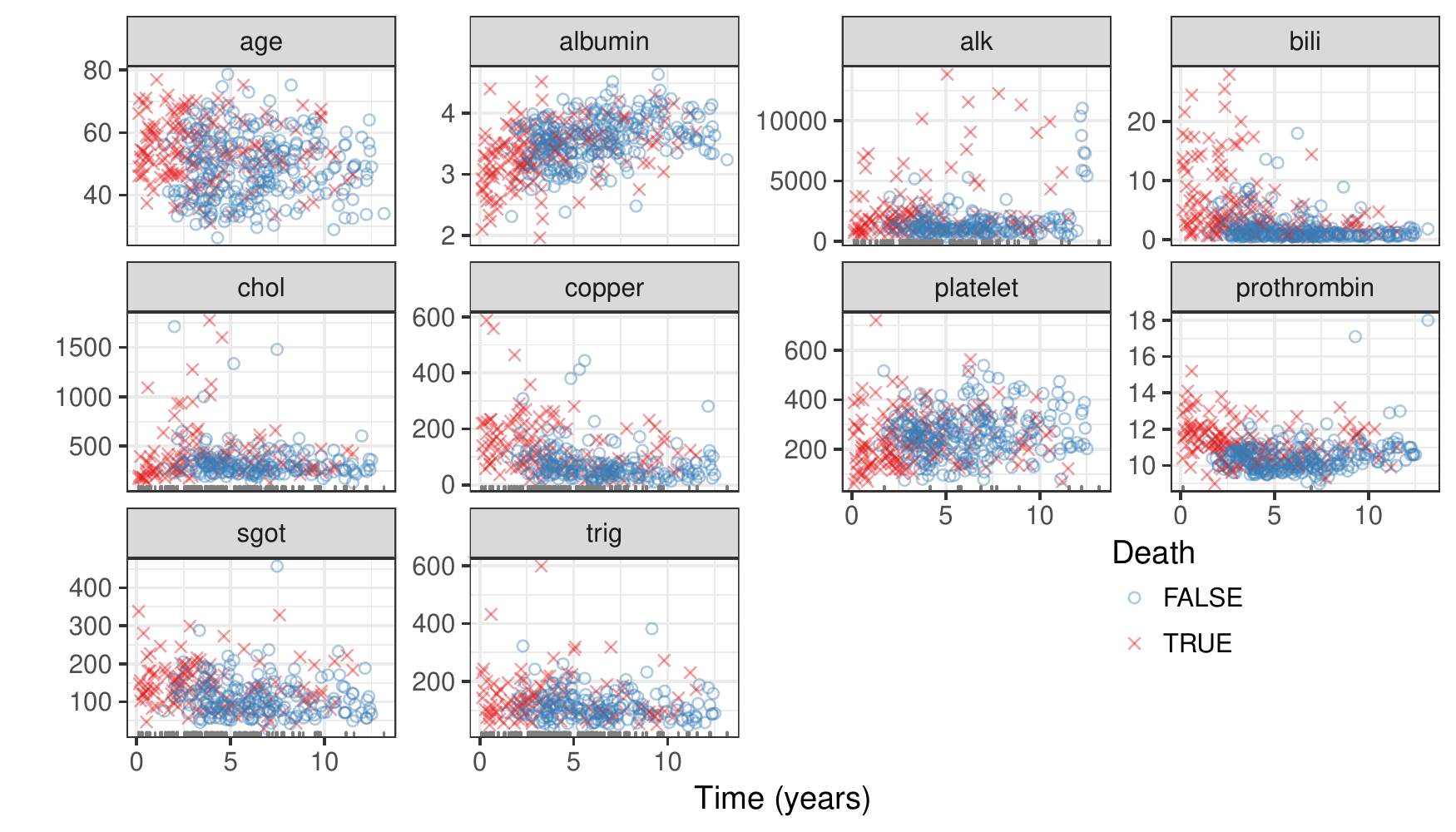}

}

\caption[EDA plots for continuous variables]{EDA plots for continuous variables. Symbols indicate observations with variable value on Y-axis against follow up time in years. Symbols are colored and shaped according to the death event  (`status` variable). Missing values are indicated by rug marks along the X-axis}\label{fig:continuousEDA}
\end{figure}
\end{Schunk}

In continuous data EDA plots (\autoref{fig:continuousEDA}), we are
looking for missingness (rug marks) and extreme or non-physical values.
For survival settings, we color and shape the points as red `x's to
indicate events, and blue circles to indicate censored observation.

Extreme value examples are evident in a few of the variables in
\autoref{fig:continuousEDA}. We are typically looking for values that
are outside of the biological range. This is often caused by
measurements recorded in differing units, which can sometimes be
corrected algorithmically. Since we can not ask the original
investigator to clarify these values in this particular study, we will
continue without modifying the data.

\begin{table}

\caption{\label{tab:missing}Missing value counts in `pbc` data set and pbc clinical trial observations (`pbc.trial`).\label{T:missing}}
\centering
\begin{tabular}[t]{lrr}
\toprule
  & pbc & pbc.trial\\
\midrule
treatment & 106 & 0\\
ascites & 106 & 0\\
hepatom & 106 & 0\\
spiders & 106 & 0\\
chol & 134 & 28\\
\addlinespace
copper & 108 & 2\\
alk & 106 & 0\\
sgot & 106 & 0\\
trig & 136 & 30\\
platelet & 11 & 4\\
\addlinespace
prothrombin & 2 & 0\\
stage & 6 & 0\\
\bottomrule
\end{tabular}
\end{table}

Both EDA figures indicate the \texttt{pbc} data set contains quite a bit
of missing data. \autoref{T:missing} shows the number of missing values
in each variable of the \texttt{pbc} data set. Of the 19 variables in
the data, 12 have missing values. The \texttt{pbc} column details
variables with missing data in the full \texttt{pbc} data set, though
there are patients that were not randomized into the trial. If we
restrict the data to the trial only, most of the missing values are also
removed, leaving only 4 variables with missing values. Therefore, we
will focus on the 312 observations from the clinical trial for the
remainder of this document. We will discuss how \pkg{randomForestSRC}
handles missing values in \autoref{random-forest-imputation}.

\subsection{PBC Model Summary}\label{pbc-model-summary}

We conclude the data set investigation with a summary
of\citep{fleming:1991} model results from Chapter 4.4. We start by
generating Kaplan--Meier (KM) survival estimates comparing the treatment
groups of DPCA and placebo. We use the \pkg{ggRandomForests}
\texttt{gg\_survival} function to generate these estimates from the data
set as follows.

\begin{Schunk}
\begin{Sinput}
R> # Create the trial and test data sets.
R> pbc.trial <- pbc %>% filter(!is.na(treatment))
R> pbc.test <- pbc %>% filter(is.na(treatment))
R>
R> # Create the gg_survival object
R> gg_dta <- gg_survival(interval = "years",
R+                       censor = "status",
R+                       by = "treatment",
R+                       data = pbc.trial,
R+                       conf.int = 0.95)
\end{Sinput}
\end{Schunk}

The code block reduces the \texttt{pbc} data set to the
\texttt{pbc.trial} which only include observations from the clinical
trial. The remaining observations are stored in the \texttt{pbc.test}
data set for later use. The \pkg{ggRandomForests} package is designed to
use a two step process in figure generation. The first step is data
generation, where we store a \texttt{gg\_survival} data object in the
\texttt{gg\_dta} object. The \texttt{gg\_survival} function uses the
\texttt{data} set, follow up \texttt{interval}, \texttt{censor}
indicator and an optional grouping argument (\texttt{by}). By default
\texttt{gg\_survival} also calculates \(95\%\) confidence band, which we
can control with the \texttt{conf.int} argument.

In the figure generation step, we use the \pkg{ggRandomForests} plot
routine \texttt{plot.gg\_survival} as shown in the following code block.
The \texttt{plot.gg\_survival} function uses the \texttt{gg\_dta} data
object to plot the survival estimate curves for each group and
corresponding confidence interval ribbons. We have used additional
\pkg{ggplot2} commands to modify the axis and legend labels
(\texttt{labs}), the legend location (\texttt{theme}) and control the
plot range of the y-axis (\texttt{coord\_cartesian}) for this figure.

\begin{Schunk}
\begin{Sinput}
R> plot(gg_dta) +
R+   labs(y = "Survival Probability", x = "Observation Time (years)",
R+        color = "Treatment", fill = "Treatment") +
R+   theme(legend.position = c(0.2, 0.2)) +
R+   coord_cartesian(y = c(0, 1.01))
\end{Sinput}
\begin{figure}[!htb]

{\centering \includegraphics{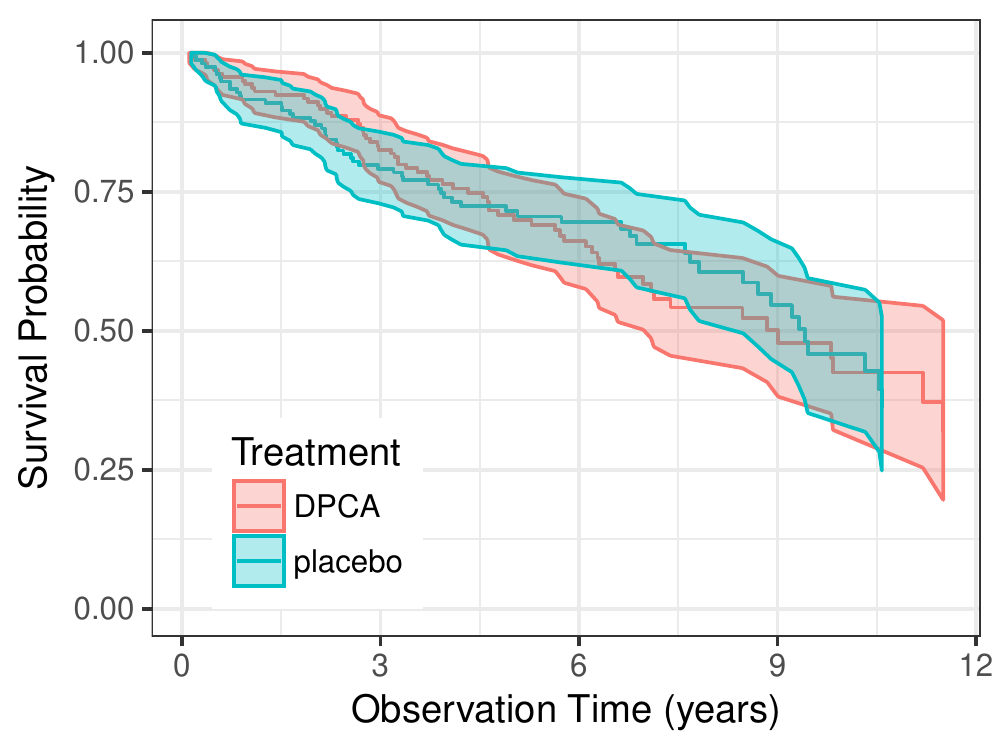}

}

\caption[Kaplan--Meier survival estimates comparing the DPCA treatment (red) with placebo (blue) groups for the pbc.trail data set]{Kaplan--Meier survival estimates comparing the DPCA treatment (red) with placebo (blue) groups for the pbc.trail data set. Median survival with shaded 95\% confidence band.}\label{fig:plot_gg_survival}
\end{figure}
\end{Schunk}

The \texttt{gg\_survival} plot of \autoref{fig:plot_gg_survival} is
analogous to\citep{fleming:1991} Figure 0.2.3 and Figure 4.4.1, showing
there is little difference between the treatment and control groups.

The \texttt{gg\_survival} function generates a variety of time-to-event
estimates, including the cumulative hazard. The follow code block
creates a cumulative hazard plot \cite[Figure 0.2.1]{fleming:1991} in
\autoref{fig:plot_gg_cum_hazard} using the same data object generated by
the original \texttt{gg\_survival} function call. The red \texttt{DPCA}
line is directly comparable to Figure 0.2.1, we've add the cumulative
hazard estimates for the \texttt{placebo} population in blue.

\begin{Schunk}
\begin{Sinput}
R> plot(gg_dta, type = "cum_haz") +
R+   labs(y = "Cumulative Hazard", x = "Observation Time (years)",
R+        color = "Treatment", fill = "Treatment") +
R+   theme(legend.position = c(0.2, 0.8)) +
R+   coord_cartesian(ylim = c(-0.02, 1.22))
\end{Sinput}
\begin{figure}[!htb]

{\centering \includegraphics{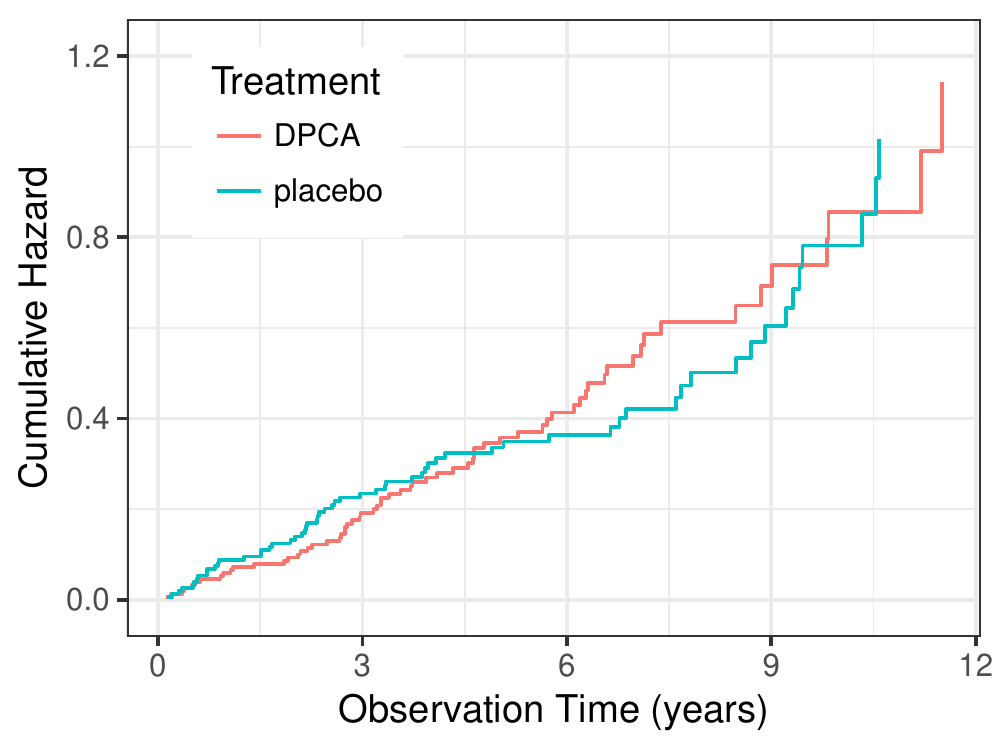}

}

\caption[Kaplan--Meier cumulative hazard estimates comparing the DPCA treatment (red) with placebo (blue) groups for the pbc data set]{Kaplan--Meier cumulative hazard estimates comparing the DPCA treatment (red) with placebo (blue) groups for the pbc data set.}\label{fig:plot_gg_cum_hazard}
\end{figure}
\end{Schunk}

In \autoref{fig:plot_gg_survival}, we demonstrated grouping on the
categorical variable (\texttt{treatment}). To demonstrate plotting
grouped survival on a continuous variable, we examine KM estimates of
survival within stratified groups of bilirubin measures. The groupings
are obtained directly from\citep{fleming:1991} Figure 4.4.2, where they
presented univariate model results of predicting survival on a function
of bilirubin.

We set up the \texttt{bili} groups on a temporary data set
(\texttt{pbc.bili}) using the \texttt{cut} function with intervals
matching the reference figure. For this example we combine the data
generation and plot steps into a single line of code. The \texttt{error}
argument of the \texttt{plot.gg\_survival} function is used to control
display of the confidence bands. We suppress the intervals for this
figure with \texttt{error\ =\ "none"} and again modify the plot display
with \pkg{ggplot2} commands to generate \autoref{fig:gg_survival-bili}.

\begin{Schunk}
\begin{Sinput}
R> pbc.bili <- pbc.trial
R> pbc.bili$bili_grp <- cut(pbc.bili$bili, breaks = c(0, 0.8, 1.3, 3.4, 29))
R>
R> plot(gg_survival(interval = "years", censor = "status", by = "bili_grp",
R+                  data = pbc.bili), error = "none") +
R+   labs(y = "Survival Probability", x = "Observation Time (years)",
R+        color = "Bilirubin")
\end{Sinput}
\begin{figure}[!htb]

{\centering \includegraphics{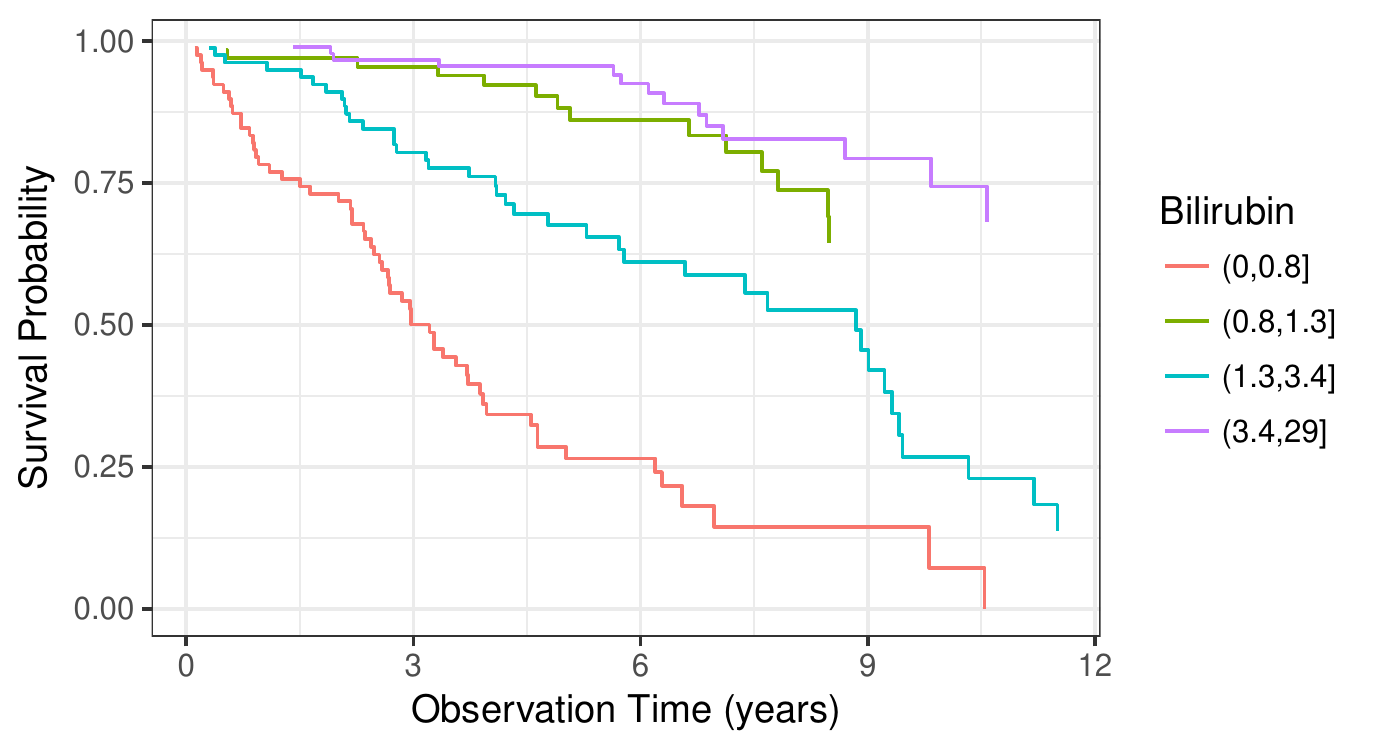}

}

\caption[Kaplan--Meier survival estimates comparing different groups of Bilirubin measures (bili) for the pbc data set]{Kaplan--Meier survival estimates comparing different groups of Bilirubin measures (bili) for the pbc data set. Groups defined in Chapter 4 of [@fleming:1991].}\label{fig:gg_survival-bili}
\end{figure}
\end{Schunk}

In Chapter 4,\citep{fleming:1991} use partial likelihood methods to
build a linear model with log transformations on some variables. We
summarize the final, biologically reasonable model in
\autoref{T:FHmodel} for later comparison with our random forest results.

\begin{table}

\caption{\label{tab:xtab}`pbc` proportional hazards model summary of 312 randomized cases in `pbc.trial` data set.  (Table 4.4.3c [@fleming:1991])\label{T:FHmodel}}
\centering
\begin{tabular}[t]{l|r|r|r}
\hline
  & Coef. & Std. Err. & Z stat.\\
\hline
Age & 0.033 & 0.009 & 3.84\\
\hline
log(Albumin) & -3.055 & 0.724 & -4.22\\
\hline
log(Bilirubin) & 0.879 & 0.099 & 8.90\\
\hline
Edema & 0.785 & 0.299 & 2.62\\
\hline
log(Prothrombin Time) & 3.016 & 1.024 & 2.95\\
\hline
\end{tabular}
\end{table}

\section{Random survival forest}\label{random-survival-forest}

A Random Forest \citep{Breiman:2001} is grown by \emph{bagging}
\citep{Breiman:1996} a collection of
\emph{classification and regression trees} (CART) \citep{cart:1984}. The
method uses a set of \(B\) \emph{bootstrap} \citep{bootstrap:1994}
samples, growing an independent tree model on each sub-sample of the
population. Each tree is grown by recursively partitioning the
population based on optimization of a \emph{split rule} over the
\(p\)-dimensional covariate space. At each split, a subset of
\(m \le p\) candidate variables are tested for the split rule
optimization, dividing each node into two daughter nodes. Each daughter
node is then split again until the process reaches the
\emph{stopping criteria} of either \emph{node purity} or
\emph{node member size}, which defines the set of
\emph{terminal (unsplit) nodes} for the tree. In regression trees, node
impurity is measured by mean squared error, whereas in classification
problems, the Gini index is used \citep{Friedman:2000}.

Random forest sorts each training set observation into one unique
terminal node per tree. Tree estimates for each observation are
constructed at each terminal node, among the terminal node members. The
Random Forest estimate for each observation is then calculated by
aggregating, averaging (regression) or votes (classification), the
terminal node results across the collection of \(B\) trees.

Random Survival Forest \citep{Ishwaran:2007, Ishwaran:2008} (RSF) are an
extension of Random Forest to analyze right censored, time to event
data. A forest of survival trees is grown using a log-rank splitting
rule to select the optimal candidate variables. Survival estimate for
each observation are constructed with a Kaplan--Meier (KM) estimator
within each terminal node, at each event time.

Random Survival Forests adaptively discover nonlinear effects and
interactions and are fully nonparametric. Averaging over many trees
enables RSF to approximate complex survival functions, including
non-proportional hazards, while maintaining low prediction error.
\citep{Ishwaran:2010a} showed that RSF is uniformly consistent and that
survival forests have a uniform approximating property in finite-sample
settings, a property not possessed by individual survival trees.

The \pkg{randomForestSRC} \texttt{rfsrc} function call grows the forest,
determining the type of forest by the response supplied in the
\texttt{formula} argument. In the following code block, we grow a random
forest for survival, by passing a survival (\texttt{Surv}) object to the
forest. The forest uses all remaining variables in the
\texttt{pbc.trial} data set to generate the RSF survival model.

\begin{Schunk}
\begin{Sinput}
R> rfsrc_pbc <- rfsrc(Surv(years, status) ~ ., data = pbc.trial,
R+                    nsplit = 10, na.action = "na.impute",
R+                    tree.err = TRUE,importance = TRUE)
\end{Sinput}
\end{Schunk}

The \texttt{print.rfsrc} function returns information on how the random
forest was grown. Here the \texttt{family\ =\ "surv"} forest has
\texttt{ntree\ =\ 1000} trees (the default \texttt{ntree} argument). The
forest selected from \texttt{ceil}\((\sqrt{p=17}) = 5\) randomly
selected candidate variables for splitting at each node, stopping when a
terminal node contained three or fewer observations. For continuous
variables, we used a random logrank split rule, which randomly selects
from \texttt{nsplit\ =\ 10} split point values, instead of optimizing
over all possible values.

\subsection{Generalization error}\label{generalization-error}

One advantage of random forest is a built in generalization error
estimate. Each bootstrap sample selects approximately \(63.2\%\) of the
population on average. The remaining \(36.8\%\) of observations, the
Out-of-Bag \citep{BreimanOOB:1996e} (OOB) sample, can be used as a hold
out test set for each tree. An OOB prediction error estimate can be
calculated for each observation by predicting the response over the set
of trees which were not trained with that particular observation.
Out-of-Bag prediction error estimates have been shown to be nearly
identical to \(n\)--fold cross validation estimates
\citep{StatisticalLearning:2009}. This feature of random forest allows
us to obtain both model fit and validation in one pass of the algorithm.

The \texttt{gg\_error} function operates on the random forest
(\texttt{rfsrc\_pbc}) object to extract the error estimates as a
function of the number of trees in the forest. The following code block
first creates a \texttt{gg\_error} data object, then uses the
\texttt{plot.gg\_error} function to create a \texttt{ggplot} object for
display in a single line of code.

\begin{Schunk}
\begin{Sinput}
R> plot(gg_error(rfsrc_pbc))
\end{Sinput}
\begin{figure}[!htb]

{\centering \includegraphics{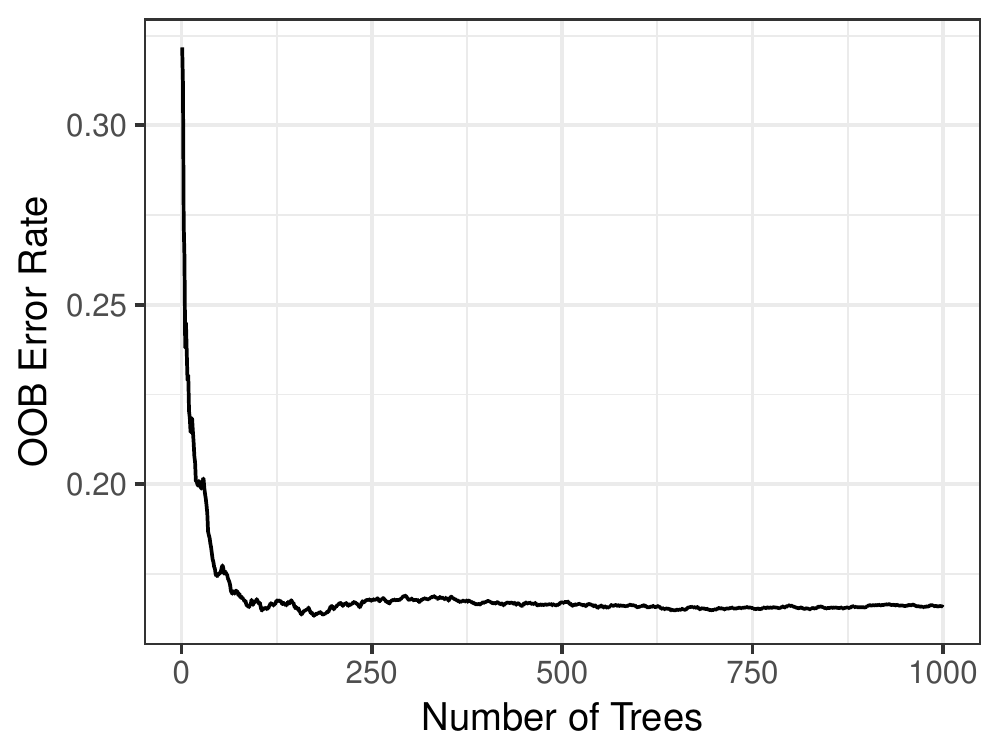}

}

\caption[Random forest OOB prediction error estimates as a function of the number of trees in the forest]{Random forest OOB prediction error estimates as a function of the number of trees in the forest.}\label{fig:errorPlot}
\end{figure}
\end{Schunk}

The \texttt{gg\_error} plot of \autoref{fig:errorPlot} demonstrates that
it does not take a large number of trees to stabilize the forest
prediction error estimate. However, to ensure that each variable has
enough of a chance to be included in the forest prediction process, we
do want to create a rather large random forest of trees.

\subsection{Training Set Prediction}\label{training-set-prediction}

The \texttt{gg\_rfsrc} function extracts the OOB prediction estimates
from the random forest. This code block executes the data extraction and
plotting in one line, since we are not interested in holding the
prediction estimates for later reuse. Each of the \pkg{ggRandomForests}
plot commands return \texttt{ggplot} objects, which we can also store
for modification or reuse later in the analysis (\texttt{ggRFsrc}
object). Note that we again use additional \pkg{ggplot2} commands to
modify the display of the plot object.

\begin{Schunk}
\begin{Sinput}
R> ggRFsrc <- plot(gg_rfsrc(rfsrc_pbc), alpha = 0.2) +
R+   scale_color_manual(values = strCol) +
R+   theme(legend.position = "none") +
R+   labs(y = "Survival Probability", x = "Time (years)") +
R+   coord_cartesian(ylim = c(-0.01, 1.01))
R> show(ggRFsrc)
\end{Sinput}
\begin{figure}[!htb]

{\centering \includegraphics{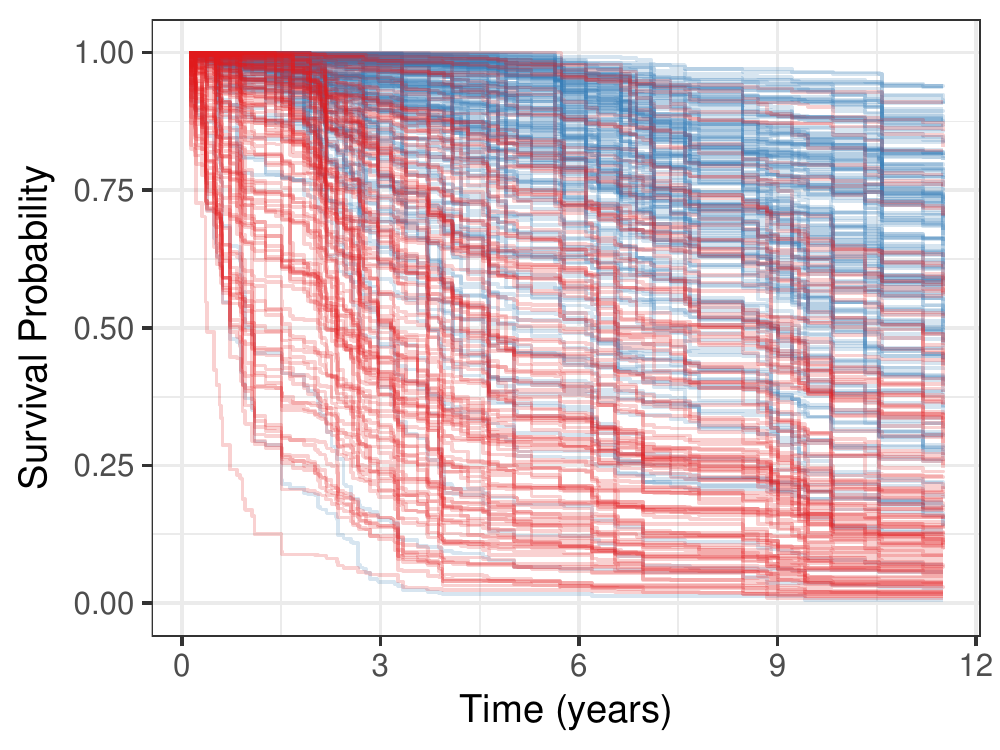}

}

\caption[Random forest OOB predicted survival]{Random forest OOB predicted survival. Blue curves correspond to censored observations, red curves correspond to observations experiencing death events.}\label{fig:rfsrc-plot}
\end{figure}
\end{Schunk}

The \texttt{gg\_rfsrc} plot of \autoref{fig:rfsrc-plot} shows the
predicted survival from our RSF model. Each line represents a single
patient in the training data set, where censored patients are colored
blue, and patients who have experienced the event (death) are colored in
red. We extend all predicted survival curves to the longest follow up
time (12 years), regardless of the actual length of a patient's follow
up time.

Interpretation of general survival properties from
\autoref{fig:rfsrc-plot} is difficult because of the number of curves
displayed. To get more interpretable results, it is preferable to plot a
summary of the survival results. The following code block compares the
predicted survival between treatment groups, as we did in
\autoref{fig:plot_gg_survival}.

\begin{Schunk}
\begin{Sinput}
R> plot(gg_rfsrc(rfsrc_pbc, by = "treatment")) +
R+   theme(legend.position = c(0.2, 0.2)) +
R+   labs(y = "Survival Probability", x = "Time (years)") +
R+   coord_cartesian(ylim = c(-0.01, 1.01))
\end{Sinput}
\begin{figure}[!htb]

{\centering \includegraphics{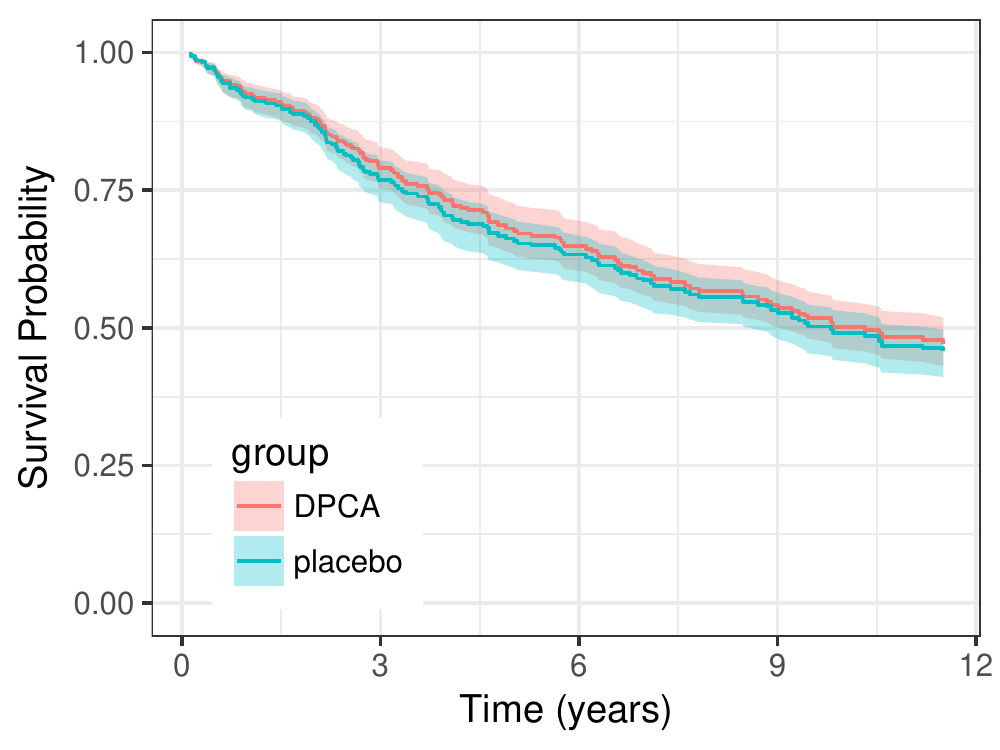}

}

\caption[Random forest predicted survival stratified by treatment groups]{Random forest predicted survival stratified by treatment groups. DPCA group in red, placebo in blue with shaded 95\% confidence bands.}\label{fig:rfsrc-mean2}
\end{figure}
\end{Schunk}

The \texttt{gg\_rfsrc} plot of \autoref{fig:rfsrc-mean2} shows the
median survival with a \(95\%\) shaded confidence band for the
\texttt{DPCA} group in red, and the \texttt{placebo} group in blue. When
calling \texttt{gg\_rfsrc} with either a \texttt{by} argument or a
\texttt{conf.int} argument, the function calculates a bootstrap
confidence interval around the median survival line. By default, the
function will calculate the \texttt{conf.int=0.95} confidence interval,
with the number of \texttt{bs.samples} equal to the number of
observations.

\subsection{Random forest imputation}\label{random-forest-imputation}

There are two modeling issues when dealing with missing data values:
\texttt{How\ does\ the\ algorithm\ build\ a\ model\ when\ values\ are\ missing\ from\ the\ training\ data?},
and
\texttt{How\ does\ the\ algorithm\ predict\ a\ response\ when\ values\ are\ missing\ from\ the\ test\ data?}.
The standard procedure for linear models is to either remove or impute
the missing data values before modelling. Removing the missingness is
done by either removing the variable with missing values (column wise)
or removing the observations (row wise). Removal is a simple solution,
but may bias results when either observations or variables are scarce.

The \pkg{randomForestSRC} package imputes missing values using
\emph{adaptive tree imputation} \citep{Ishwaran:2008}. Rather than
impute missing values before growing the forest, the algorithm takes a
\texttt{just-in-time} approach. At each node split, the set of
\texttt{mtry} candidate variables is checked for missing values. Missing
values are then imputed by randomly drawing values from non-missing data
within the node. The split-statistic is then calculated on observations
that were not missing values. The imputed values are used to sort
observations into the subsequent daughter nodes and then discarded
before the next split occurs. The process is repeated until the stopping
criteria is reached and all observations are sorted into terminal nodes.

A final imputation step can be used to fill in missing values from
within the terminal nodes. This step uses a process similar to the
previous imputation but uses the OOB non-missing terminal node data for
the random draws. These values are aggregated (averaging for continuous
variables, voting for categorical variables) over the \texttt{ntree}
trees in the forest to estimate an imputed data set. By default, the
missing values are not filled into the training data, but are available
within the forest object for later use if desired.

Adaptive tree imputation still requires the missing at random
assumptions \citep{Rubin:1976}. At each imputation step, the random
forest assumes that similar observations are grouped together within
each node. The random draws used to fill in missing data do not bias the
split rule, but only sort observations similar in non-missing data into
like nodes. An additional feature of this approach is the ability of
predicting on test set observations with missing values.

\subsection{Test set predictions}\label{test-set-predictions}

The strength of adaptive tree imputation becomes clear when doing
prediction on test set observations. If we want to predict survival for
patients that did not participate in the trial using the model we
created in Section \ref{random-survival-forest}, we need to somehow
account for the missing values detailed in \autoref{T:missing}.

The \texttt{predict.rfsrc} call takes the forest object
(\texttt{rfsrc\_pbc}), and the test data set (\texttt{pbc\_test}) and
returns a predicted survival using the same forest imputation method for
missing values within the test data set
(\texttt{na.action="na.impute"}).

\begin{Schunk}
\begin{Sinput}
R> rfsrc_pbc_test <- predict(rfsrc_pbc, newdata = pbc.test,
R+                           na.action = "na.impute",
R+                           importance = TRUE)
\end{Sinput}
\end{Schunk}

The forest summary indicates there are 106 test set observations with 36
deaths and the predicted error rate is \(19.1\%\). We plot the predicted
survival just as we did the training set estimates.

\begin{Schunk}
\begin{Sinput}
R> plot(gg_rfsrc(rfsrc_pbc_test), alpha=.2) +
R+   scale_color_manual(values = strCol) +
R+   theme(legend.position = "none") +
R+   labs(y = "Survival Probability", x = "Time (years)") +
R+   coord_cartesian(ylim = c(-0.01, 1.01))
\end{Sinput}
\begin{figure}[!htb]

{\centering \includegraphics{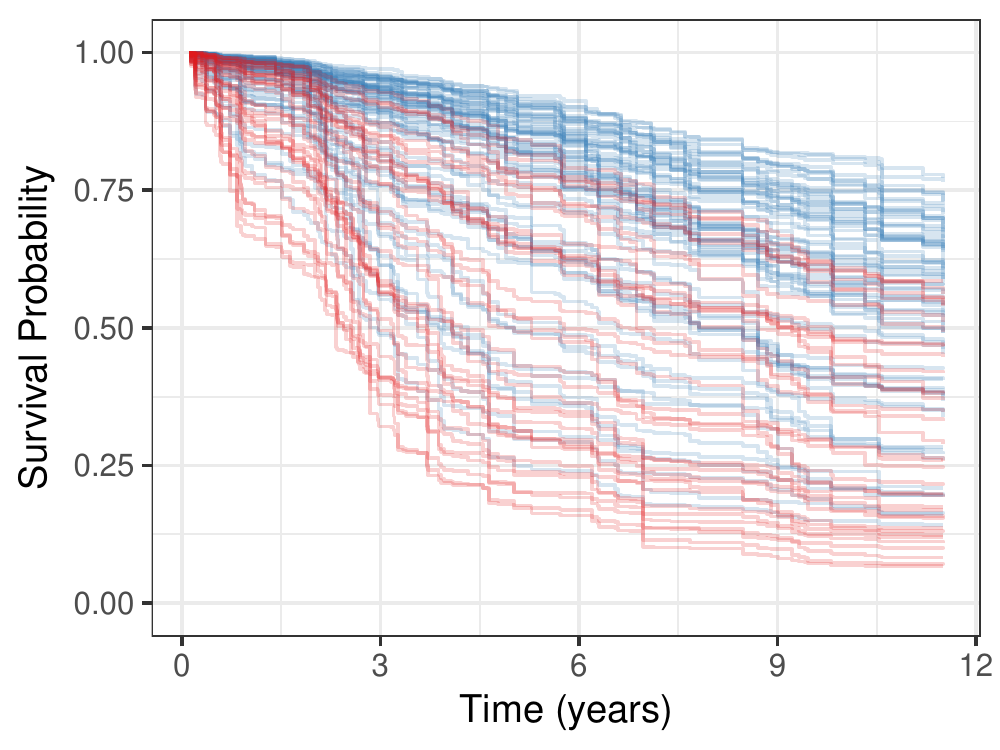}

}

\caption[Random forest survival estimates for patients in the pbc.test data set]{Random forest survival estimates for patients in the pbc.test data set. Blue curves correspond to censored patients, red curves correspond to patients experiencing a death event.}\label{fig:predictPlot}
\end{figure}
\end{Schunk}

The \texttt{gg\_rfsrc} plot of \autoref{fig:predictPlot} shows the test
set predictions, similar to the training set predictions in
\autoref{fig:rfsrc-plot}, though with fewer patients the survival curves
do not cover the same area of the figure. It is important to note that
because \autoref{fig:rfsrc-plot} is constructed with OOB estimates, the
survival results are comparable as estimates from unseen observations in
\autoref{fig:predictPlot}.

\section{Variable selection}\label{variable-selection}

Random forest is not a parsimonious method, but uses all variables
available in the data set to construct the response predictor. Also,
unlike parametric models, random forest does not require the explicit
specification of the functional form of covariates to the response.
Therefore there is no explicit \(p\)-value/significance test for
variable selection with a random forest model. Instead, RF ascertains
which variables contribute to the prediction through the split rule
optimization, optimally choosing variables which separate observations.

The typical goal of a random forest analysis is to build a
\emph{prediction} model, in contrast to extracting \emph{information}
regarding the underlying process \citep{Breiman:twoCultures:2001}. There
is not usually much care given in how variables are included into the
training data set. Since the goal is prediction, investigators often
include the ``kitchen sink'' if it can help.

In contrast, in survival settings we are typically also interested in
how we can possibly improve the the outcome of interest. To achieve
this, for understandable inference, it is important to avoid both
duplication and transformations of variables whenever possible when
building our data sets. Duplication of variables, including multiple
measures of a similar covariate, can reduce or mask the importance of
the covariate. Transformations can also mask importance as well as make
interpretation of the inference results difficult to impossible.

In this Section, We explore two separate approaches to investigate the
RF variable selection process. Variable Importance
(\autoref{variable-importance}), a property related to variable
misspecification, and Minimal Depth (\autoref{minimal-depth}), a
property derived from the construction of the trees within the forest.

\subsection{Variable Importance}\label{variable-importance}

\emph{Variable importance} (VIMP) was originally defined in CART using a
measure involving surrogate variables (see Chapter 5 of
\citep{cart:1984}). The most popular VIMP method uses a prediction error
approach involving ``noising-up'' each variable in turn. VIMP for a
variable \(x_v\) is the difference between prediction error when \(x_v\)
is randomly permuted, compared to prediction error under the observed
values \citep{Breiman:2001, Liaw:2002, Ishwaran:2007, Ishwaran:2008}.

Since VIMP is the difference in OOB prediction error before and after
permutation, a large VIMP value indicates that misspecification detracts
from the predictive accuracy in the forest. VIMP close to zero indicates
the variable contributes nothing to predictive accuracy, and negative
values indicate the predictive accuracy \emph{improves} when the
variable is misspecified. In the later case, we assume noise is more
informative than the true variable. As such, we ignore variables with
negative and near zero values of VIMP, relying on large positive values
to indicate that the predictive power of the forest is dependent on
those variables.

The \texttt{gg\_vimp} function extracts VIMP measures for each of the
variables used to grow the forest. The \texttt{plot.gg\_vimp} function
shows the variables, in VIMP rank order, labeled with the named vector
in the \texttt{lbls} argument.

\begin{Schunk}
\begin{Sinput}
R> plot(gg_vimp(rfsrc_pbc), lbls = st.labs) +
R+   theme(legend.position = c(0.8, 0.2)) +
R+   labs(fill = "VIMP > 0")
\end{Sinput}
\begin{figure}[!htb]

{\centering \includegraphics{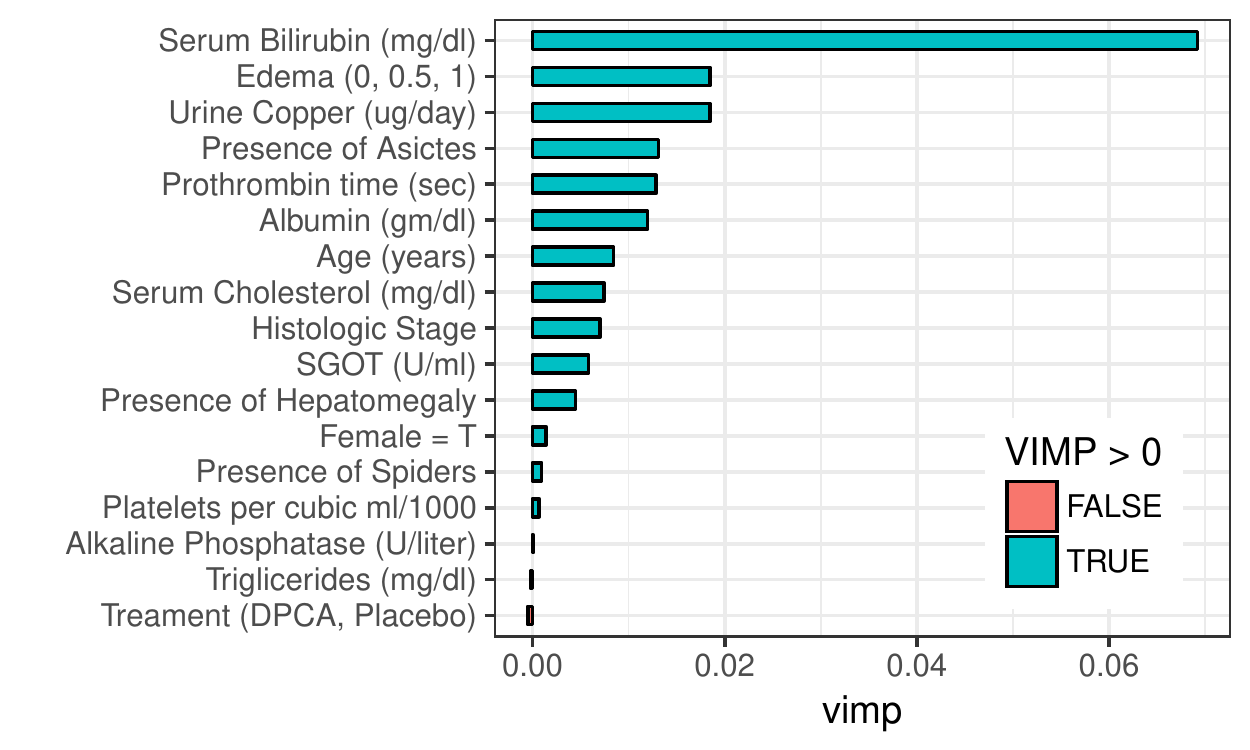}

}

\caption[Random forest Variable Importance (VIMP)]{Random forest Variable Importance (VIMP). Blue bars indicates positive VIMP, red indicates negative VIMP. Importance is relative to positive length of bars.}\label{fig:rf-vimp}
\end{figure}
\end{Schunk}

The \texttt{gg\_vimp} plot of \autoref{fig:rf-vimp} details VIMP ranking
for the \texttt{pbc.trial} baseline variables, from the largest (Serum
Bilirubin) at the top, to smallest (Treament (DPCA, Placebo)) at the
bottom. VIMP measures are shown using bars to compare the scale of the
error increase under permutation and colored by the sign of the measure
(red for negative values). Note that four of the five highest ranking
variables by VIMP match those selected by the\citep{fleming:1991} model
listed in \autoref{T:FHmodel}, with urine copper (2) ranking higher than
age (8). We will return to this in
\autoref{variable-selection-comparison}.

\subsection{Minimal Depth}\label{minimal-depth}

In VIMP, prognostic risk factors are determined by testing the forest
prediction under alternative data settings, ranking the most important
variables according to their impact on predictive ability of the forest.
An alternative method uses inspection of the forest construction to rank
variables. \emph{Minimal depth} \citep{Ishwaran:2010, Ishwaran:2011}
assumes that variables with high impact on the prediction are those that
most frequently split nodes nearest to the root node, where they
partition the largest samples of the population.

Within each tree, node levels are numbered based on their relative
distance to the root of the tree (with the root at 0). Minimal depth
measures important risk factors by averaging the depth of the first
split for each variable over all trees within the forest. The assumption
in the metric is that smaller minimal depth values indicate the variable
separates large groups of observations, and therefore has a large impact
on the forest prediction.

In general, to select variables according to VIMP, we examine the VIMP
values, looking for some point along the ranking where there is a large
difference in VIMP measures. Given minimal depth is a quantitative
property of the forest construction, \cite{Ishwaran:2010} also derive an
analytic threshold for evidence of variable impact. A simple optimistic
threshold rule uses the mean of the minimal depth distribution,
classifying variables with minimal depth lower than this threshold as
important in forest prediction.

The \pkg{randomForestSRC} \texttt{var.select} function uses the minimal
depth methodology for variable selection, returning an object with both
minimal depth and vimp measures. The \pkg{ggRandomForests}
\texttt{gg\_minimal\_depth} function is analogous to the
\texttt{gg\_vimp} function. Variables are ranked from most important at
the top (minimal depth measure), to least at the bottom (maximal minimal
depth).

\begin{Schunk}
\begin{Sinput}
R> varsel_pbc <- var.select(rfsrc_pbc)
\end{Sinput}
\begin{Soutput}
minimal depth variable selection ...


-----------------------------------------------------------
family             : surv
var. selection     : Minimal Depth
conservativeness   : medium
x-weighting used?  : TRUE
dimension          : 17
sample size        : 312
ntree              : 1000
nsplit             : 10
mtry               : 5
nodesize           : 3
refitted forest    : FALSE
model size         : 14
depth threshold    : 6.7549
PE (true OOB)      : 16.502


Top variables:
            depth  vimp
bili        1.708 0.067
albumin     2.483 0.012
copper      2.904 0.015
prothrombin 2.931 0.014
chol        3.227 0.006
platelet    3.329 0.000
edema       3.333 0.016
sgot        3.677 0.007
age         3.702 0.009
alk         4.039 0.001
trig        4.514 0.000
ascites     5.194 0.013
stage       5.247 0.007
hepatom     6.476 0.003
-----------------------------------------------------------
\end{Soutput}
\begin{Sinput}
R> gg_md <- gg_minimal_depth(varsel_pbc, lbls = st.labs)
R> # print(gg_md)
\end{Sinput}
\end{Schunk}

The \texttt{gg\_minimal\_depth} summary mostly reproduces the output
from the \texttt{var.select} function from the \pkg{randomForestSRC}
package. We report the minimal depth threshold (\texttt{threshold}
6.755) and the number of variables with depth below that threshold
(\texttt{model\ size} 14). We also list a table of the top (14) selected
variables, in minimal depth rank order with the associated VIMP
measures. The minimal depth numbers indicate that \texttt{bili} tends to
split between the first and second node level, and the next three
variables (\texttt{albumin}, \texttt{copper}, \texttt{prothrombin})
split between the second and third levels on average.

\begin{Schunk}
\begin{Sinput}
R> plot(gg_md, lbls = st.labs)
\end{Sinput}
\begin{figure}[!htb]

{\centering \includegraphics{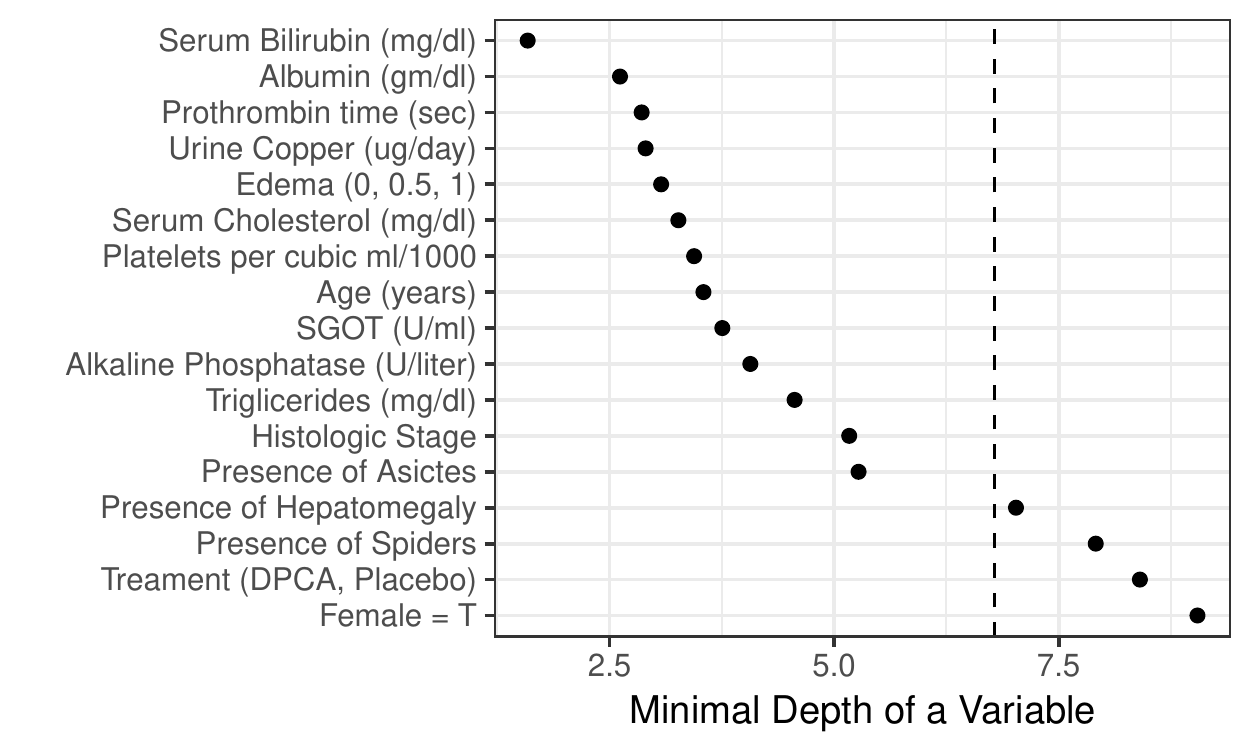}

}

\caption[Minimal Depth variable selection]{Minimal Depth variable selection. Low minimal depth indicates important variables. The dashed line is the threshold of maximum value for variable selection.}\label{fig:mindepth-plot}
\end{figure}
\end{Schunk}

The \texttt{gg\_minimal\_depth} plot of \autoref{fig:mindepth-plot} is
similar to the \texttt{gg\_vimp} plot in \autoref{fig:rf-vimp}, ranking
variables from most important at the top (minimal depth measure), to
least at the bottom (maximal minimal depth). The vertical dashed line
indicates the minimal depth threshold where smaller minimal depth values
indicate higher importance and larger values indicate lower importance.

\subsection{Variable selection
comparison}\label{variable-selection-comparison}

Since the VIMP and Minimal Depth measures use different criteria, we
expect the variable ranking to be somewhat different. We use
\texttt{gg\_minimal\_vimp} function to compare rankings between minimal
depth and VIMP in \autoref{fig:depthVimp}.

\begin{Schunk}
\begin{Sinput}
R> plot(gg_minimal_vimp(gg_md), lbls = st.labs) +
R+   theme(legend.position=c(0.8, 0.2))
\end{Sinput}
\begin{figure}[!htb]

{\centering \includegraphics{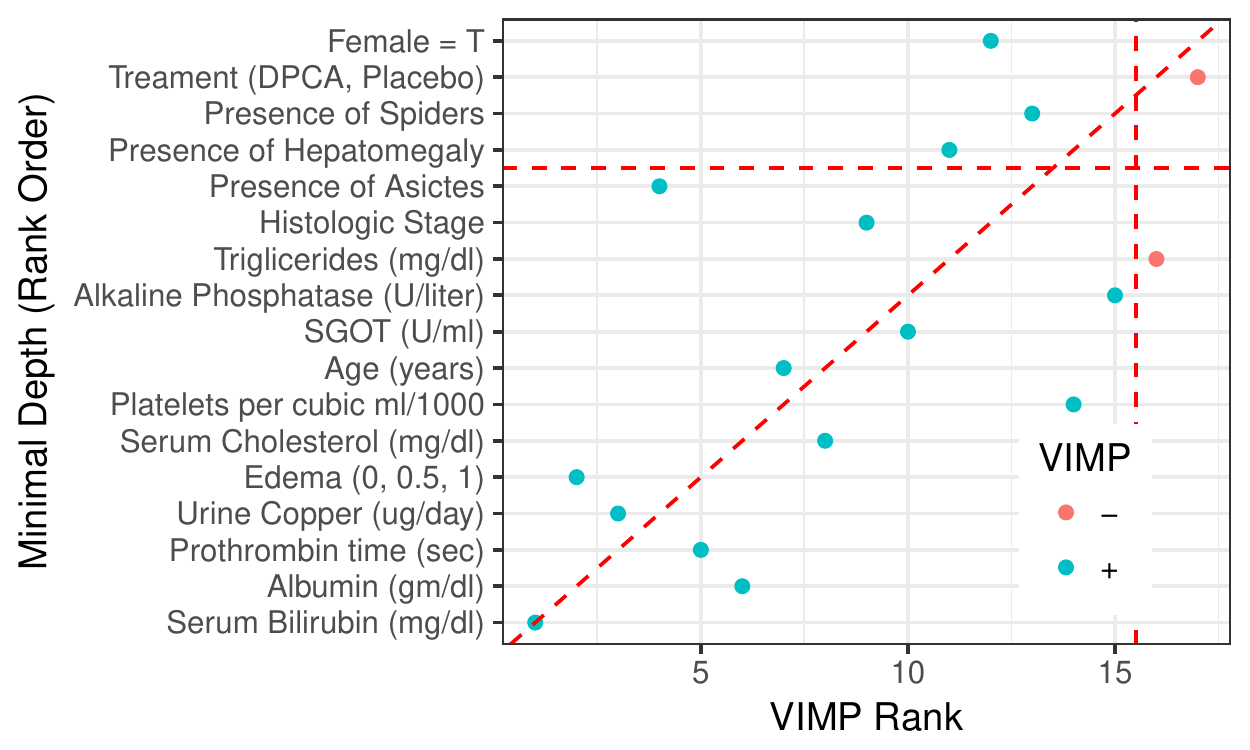}

}

\caption[Comparing Minimal Depth and Vimp rankings]{Comparing Minimal Depth and Vimp rankings. Points on the red dashed line are ranked equivalently, points above have higher VIMP ranking, those below have higher minimal depth ranking.}\label{fig:depthVimp}
\end{figure}
\end{Schunk}

The points along the red dashed line indicate where the measures are in
agreement. Points above the red dashed line are ranked higher by VIMP
than by minimal depth, indicating the variables are more sensitive to
misspecification. Those below the line have a higher minimal depth
ranking, indicating they are better at dividing large portions of the
population. The further the points are from the line, the more the
discrepancy between measures.

\begin{Schunk}
\begin{table}

\caption{\label{tab:models}Comparison of variable selection criteria. Minimal depth ranking, VIMP ranking and [@fleming:1991] (FH) proportional hazards model ranked according to `abs(Z stat)` from Table \ref{T:FHmodel}.\label{T:modelComp}}
\centering
\begin{tabular}[t]{l|r|r|r}
\hline
Variable & FH & Min depth & VIMP\\
\hline
bili & 1 & 1 & 1\\
\hline
albumin & 2 & 2 & 6\\
\hline
copper & NA & 3 & 3\\
\hline
prothrombin & 4 & 4 & 4\\
\hline
chol & NA & 5 & 10\\
\hline
platelet & NA & 6 & 16\\
\hline
edema & 5 & 7 & 2\\
\hline
sgot & NA & 8 & 8\\
\hline
age & 3 & 9 & 7\\
\hline
alk & NA & 10 & 13\\
\hline
trig & NA & 11 & 15\\
\hline
ascites & NA & 12 & 5\\
\hline
stage & NA & 13 & 9\\
\hline
hepatom & NA & 14 & 11\\
\hline
\end{tabular}
\end{table}

\end{Schunk}

We examine the ranking of the different variable selection methods
further in \autoref{T:modelComp}. We can use the Z statistic from
\autoref{T:FHmodel} to rank variables selected in
the\citep{fleming:1991} model to compare with variables selected by
minimal depth and VIMP. The table is constructed by taking the top
ranked minimal depth variables (below the selection threshold) and
matching the VIMP ranking and\citep{fleming:1991} model transforms. We
see all three methods indicate a strong relation of serum bilirubin to
survival, and overall, the minimal depth and VIMP rankings agree
reasonably well with the\citep{fleming:1991} model.

The minimal depth selection process reduced the number of variables of
interest from\textasciitilde{}17 to 14, which is still a rather large
subset of interest. An obvious selection set is to examine the five
variables selected by\citep{fleming:1991}. Combining the Minimal Depth
and\citep{fleming:1991} model, there may be evidence to keep the top 7
variables. Though minimal depth does not indicate the \texttt{edema}
variable is very interesting, VIMP ranking does agree with the
proportional hazards model, indicating we might not want to remove the
\texttt{edema} variable. Both minimal depth and VIMP suggest including
\texttt{copper}, a measure associated with liver disease.

Regarding the \texttt{chol} variable, recall missing data summary of
\autoref{T:missing}. In in the trial data set, there were 28
observations missing \texttt{chol} values. The forest imputation
randomly sorts observations with missing values into daughter nodes when
using the \texttt{chol} variable, which is also how
\pkg{randomForestSRC} calculates VIMP. We therefore expect low values
for VIMP when a variable has a reasonable number of missing values.

Restricting our remaining analysis to the five\citep{fleming:1991}
variables, plus the \texttt{copper} retains the biological sense of
these analysis. We will now examine how these six variables are related
to survival using variable dependence methods to determine the direction
of the effect and verify that the log transforms used
by\citep{fleming:1991} are appropriate.

\section{Variable/Response
dependence}\label{variableresponse-dependence}

As random forest is not parsimonious, we have used minimal depth and
VIMP to reduce the number of variables to a manageable subset. Once we
have an idea of which variables contribute most to the predictive
accuracy of the forest, we would like to know how the response depends
on these variables.

Although often characterized as a \emph{black box} method, the forest
predictor is a function of the predictor variables
\(\hat{f}_{RF} = f(x).\) We use graphical methods to examine the forest
predicted response dependency on covariates. We again have two options,
variable dependence plots (\autoref{variable-dependence}) are quick and
easy to generate, and partial dependence plots
(\autoref{partial-dependence}) are more computationally intensive but
give us a risk adjusted look at variable dependence.

\subsection{Variable dependence}\label{variable-dependence}

\emph{Variable dependence} plots show the predicted response relative to
a covariate of interest, with each training set observation represented
by a point on the plot. Interpretation of variable dependence plots can
only be in general terms, as point predictions are a function of all
covariates in that particular observation.

Variable dependence is straight forward to calculate, involving only the
getting the predicted response for each observation. In survival
settings, we must account for the additional dimension of time. We plot
the response at specific time points of interest, for example survival
at 1 or 3 years.

\begin{Schunk}
\begin{Sinput}
R> ggRFsrc + geom_vline(aes(xintercept = 1), linetype = "dashed") +
R+    geom_vline(aes(xintercept = 3), linetype = "dashed") +
R+   coord_cartesian(xlim = c(0, 5))
\end{Sinput}
\begin{figure}[!htb]

{\centering \includegraphics{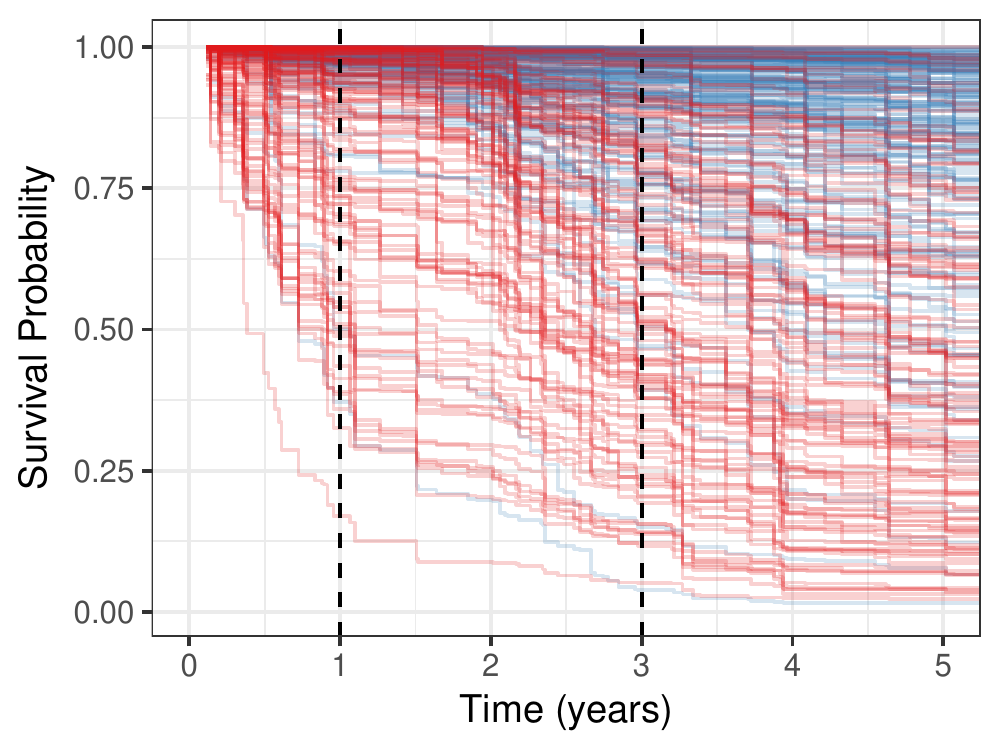}

}

\caption{Random forest predicted survival (Figure \ref{fig:rfsrc-plot}) with vertical dashed lines indicate the 1 and 3 year survival estimates.}\label{fig:rfsrc-plot3Mnth}
\end{figure}
\end{Schunk}

The \texttt{gg\_rfsrc} of \autoref{fig:rfsrc-plot3Mnth} identical to
\autoref{fig:rfsrc-plot} (stored in the \texttt{ggRFsrc} variable) with
the addition of a vertical dashed line at the 1 and 3 year survival
time. A variable dependence plot is generated from the predicted
response value of each survival curve at the intersecting time line
plotted against covariate value for that observation. This can be
visualized as taking a slice of the predicted response at each time
line, and spreading the resulting points out along the variable of
interest.

The \texttt{gg\_variable} function extracts the training set variables
and the predicted OOB response from \texttt{rfsrc} and \texttt{predict}
objects. In the following code block, we store the \texttt{gg\_variable}
data object for later use (\texttt{gg\_v}), as all remaining variable
dependence plots can be constructed from this object.

\begin{Schunk}
\begin{Sinput}
R> gg_v <- gg_variable(rfsrc_pbc, time = c(1, 3),
R+                     time.labels = c("1 Year", "3 Years"))
R>
R> plot(gg_v, xvar = "bili", alpha = 0.4) + #, se=FALSE
R+   labs(y = "Survival", x = st.labs["bili"]) +
R+   theme(legend.position = "none") +
R+   scale_color_manual(values = strCol, labels = event.labels) +
R+   scale_shape_manual(values = event.marks, labels = event.labels) +
R+   coord_cartesian(ylim = c(-0.01, 1.01))
\end{Sinput}
\begin{figure}[!htb]

{\centering \includegraphics{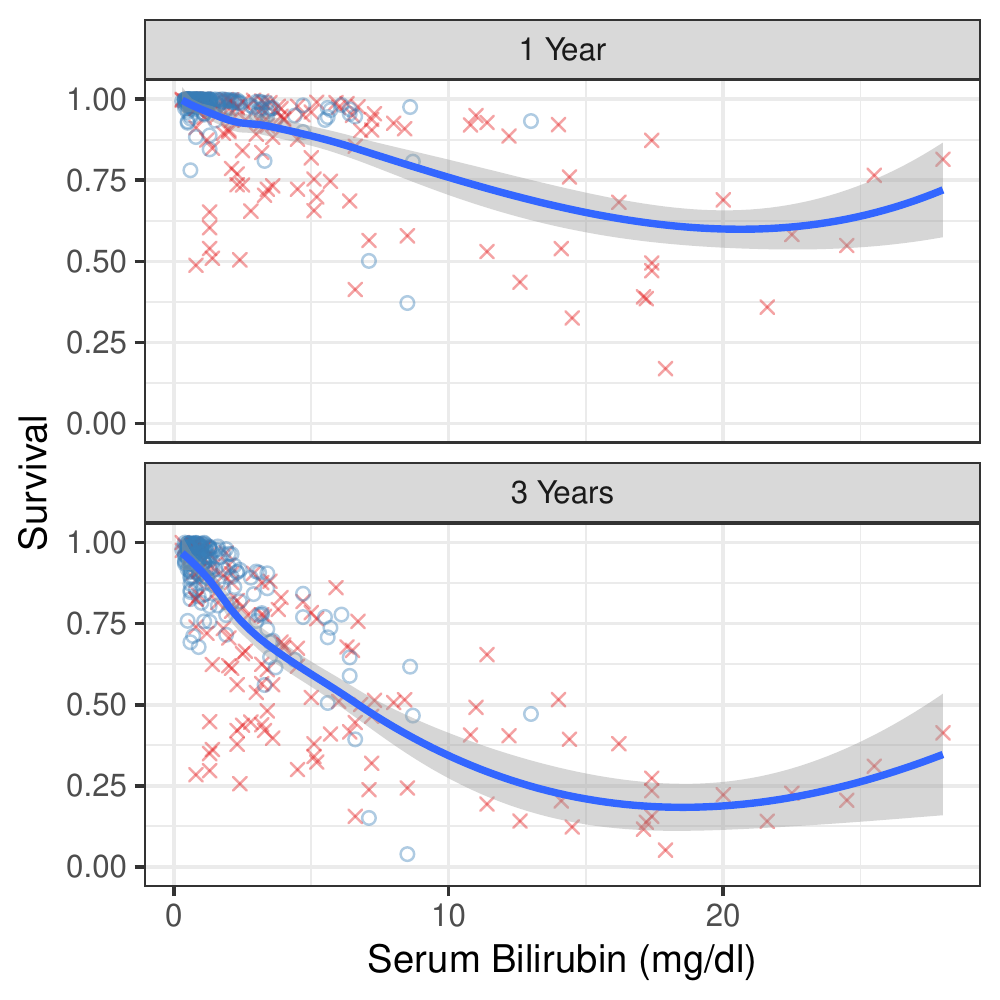}

}

\caption[Variable dependence of survival at 1 and 3 years on bili variable]{Variable dependence of survival at 1 and 3 years on bili variable. Individual cases are marked with blue circles (alive or censored) and red x (dead). Loess smooth curve with shaded 95\% confidence band indicates decreasing survival with increasing bilirubin.}\label{fig:variable-plotbili}
\end{figure}
\end{Schunk}

The \texttt{gg\_variable} plot of \autoref{fig:variable-plotbili} shows
variable dependence for the Serum Bilirubin (\texttt{bili}) variable.
Again censored cases are shown as blue circles, events are indicated by
the red \texttt{x} symbols. Each predicted point is dependent on the
full combination of all other covariates, not only on the covariate
displayed in the dependence plot. The smooth loess line
\citep{cleveland:1981, cleveland:1988} indicates the trend of the
prediction over the change in the variable.

Examination of \autoref{fig:variable-plotbili} indicates most of the
cases are grouped in the lower end of \texttt{bili} values. We also see
that most of the higher values experienced an event. The ``normal''
range of Bilirubin is from 0.3 to 1.9 mg/dL, indicating the distribution
from our population is well outside the normal range. These values make
biological sense considering Bilirubin is a pigment created in the
liver, the organ effected by the PBC disease. The figure also shows that
the risk of death increases as time progresses. The risk at 3 years is
much greater than that at 1 year for patients with high Bilirubin values
compared to those with values closer to the normal range.

The \texttt{plot.gg\_variable} function call operates on the
\texttt{gg\_variable} object controlled by the list of variables of
interest in the \texttt{xvar} argument. By default, the
\texttt{plot.gg\_variable} function returns a list of \texttt{ggplot}
objects, one figure for each variable named in \texttt{xvar}. The
remaining arguments are passed to internal \pkg{ggplot2} functions
controlling the display of the figure. The \texttt{se} argument is
passed to the internal call to \texttt{geom\_smooth} for fitting smooth
lines to the data. The \texttt{alpha} argument lightens the coloring
points in the \texttt{geom\_point} call, making it easier to see point
over plotting. We also demonstrate modification of the plot labels using
the \texttt{labs} function and point attributes with the
\texttt{scale\_} functions.

An additional \texttt{plot.gg\_variable} argument
(\texttt{panel\ =\ TRUE}) can be used to combine multiple variable
dependence plots into a single figure. In the following code block, we
plot the remaining continuous variables of interest found in
\autoref{T:modelComp}.

\begin{Schunk}
\begin{Sinput}
R> xvar <- c("bili", "albumin", "copper", "prothrombin", "age")
R> xvar.cat <- c("edema")
R>
R> plot(gg_v, xvar = xvar[-1], panel = TRUE, alpha = 0.4) +
R+   labs(y = "Survival") +
R+   theme(legend.position = "none") +
R+   scale_color_manual(values = strCol, labels = event.labels) +
R+   scale_shape_manual(values = event.marks, labels = event.labels) +
R+   coord_cartesian(ylim = c(-0.05, 1.05))
\end{Sinput}
\begin{figure}[!htb]

{\centering \includegraphics{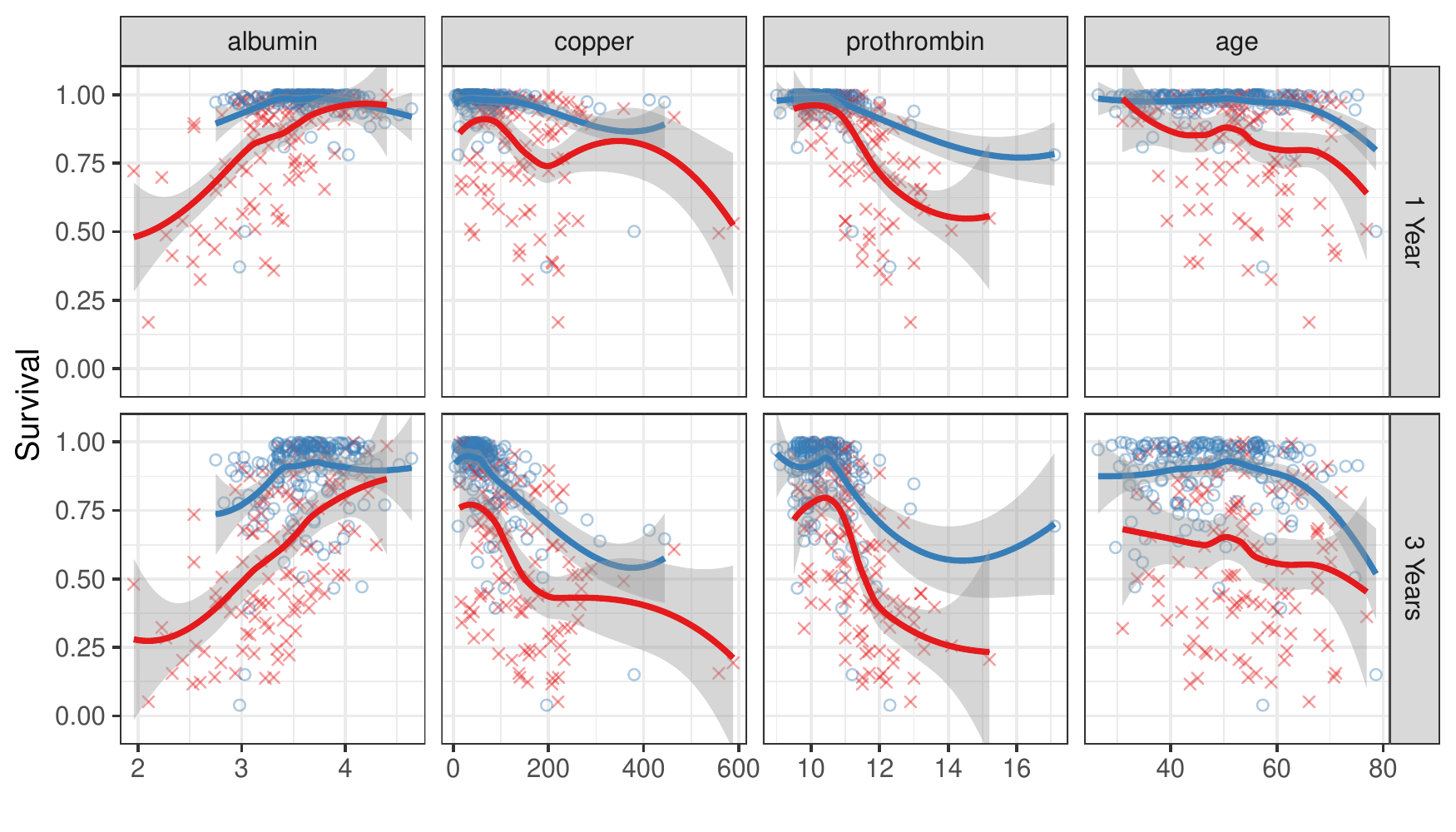}

}

\caption[Variable dependence of predicted survival at 1 and 3 years on continuous variables of interest]{Variable dependence of predicted survival at 1 and 3 years on continuous variables of interest. Individual cases are marked with blue circles for censored cases and red x for death events. Loess smooth curve indicates the survival trend with increasing values.}\label{fig:variable-plot}
\end{figure}
\end{Schunk}

The \texttt{gg\_variable} plot in \autoref{fig:variable-plot} displays a
panel of the remaining continuous variable dependence plots. The panels
are sorted in the order of variables in the \texttt{xvar} argument and
include a smooth loess line \citep{cleveland:1981, cleveland:1988} to
indicate the trend of the prediction dependence over the covariate
values. The \texttt{se=FALSE} argument turns off the loess confidence
band, and the \texttt{span=1} argument controls the degree of smoothing.

The figures indicate that survival increases with \texttt{albumin}
level, and decreases with \texttt{bili}, \texttt{copper},
\texttt{prothrombin} and \texttt{age}. Note the extreme value of
\texttt{prothrombin} (\textgreater{} 16) influences the loess curve more
than other points, which would make it a candidate for further
investigation.

We expect survival at 3 years to be lower than at 1 year. However,
comparing the two time plots for each variable does indicate a
difference in response relation for \texttt{bili}, \texttt{copper} and
\texttt{prothrombine}. The added risk for high levels of these variables
at 3 years indicates a non-proportional hazards response. The similarity
between the time curves for \texttt{albumin} and \texttt{age} indicates
the effect of these variables is constant over the disease progression.

There is not a convenient method to panel scatter plots and boxplots
together, so we recommend creating panel plots for each variable type
separately. We plot the categorical variable (\texttt{edema}) in
\autoref{fig:variable-plotCat} separately from the continuous variables
in \autoref{fig:variable-plot}.

\begin{Schunk}
\begin{Sinput}
R> plot(gg_v, xvar = xvar.cat, alpha = 0.4) + labs(y = "Survival") +
R+   theme(legend.position = "none") +
R+   scale_color_manual(values = strCol, labels = event.labels) +
R+   scale_shape_manual(values = event.marks, labels = event.labels) +
R+   coord_cartesian(ylim = c(-0.01, 1.02))
\end{Sinput}
\begin{figure}[!htb]

{\centering \includegraphics{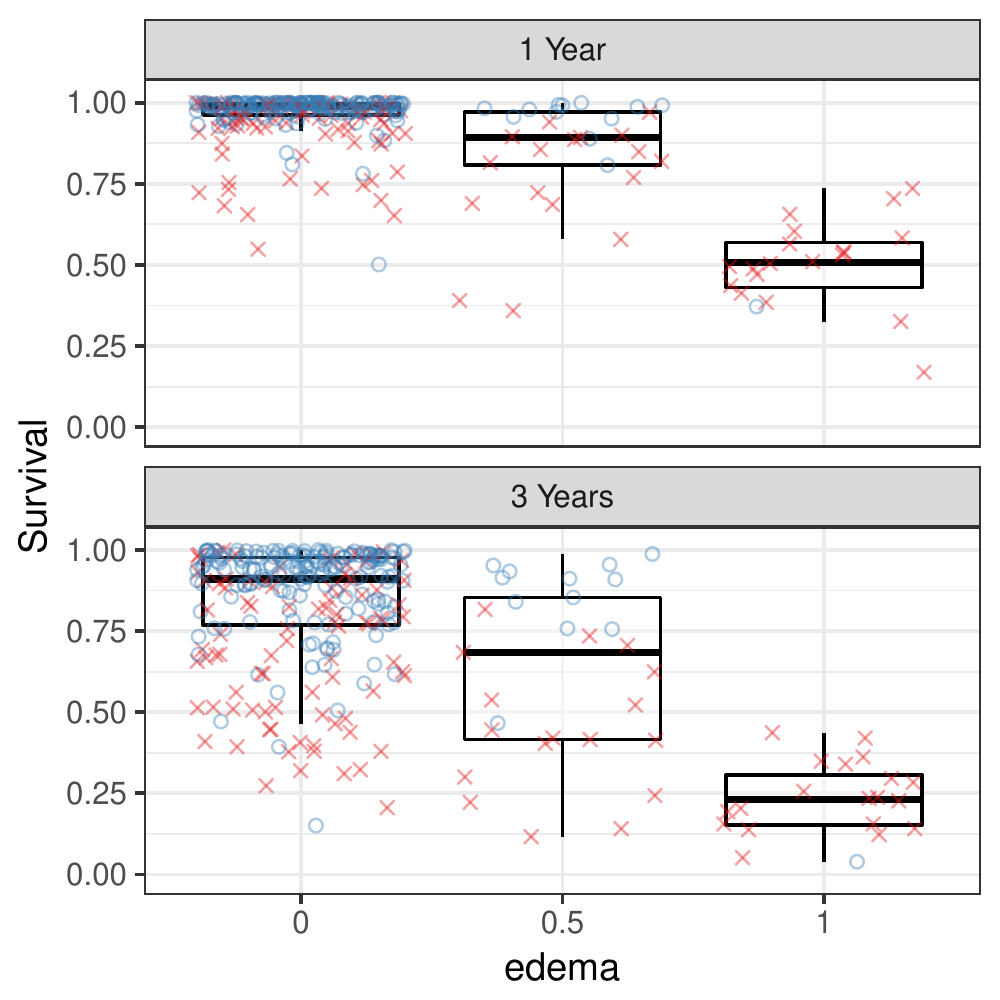}

}

\caption[Variable dependence of survival 1 and 3 years on edema categorical variable]{Variable dependence of survival 1 and 3 years on edema categorical variable. Symbols with blue circles indicate censored cases and red x indicate death events. Boxplots indicate distribution of predicted survival for all observations within each edema group.}\label{fig:variable-plotCat}
\end{figure}
\end{Schunk}

The \texttt{gg\_variable} plot of \autoref{fig:variable-plotCat} for
categorical variable dependence displays boxplots to examine the
distribution of predicted values within each level of the variable. The
points are plotted with a jitter to see the censored and event markers
more clearly. The boxes are shown with horizontal bars indicating the
median, 75th (top) and 25th (bottom) percentiles. Whiskers extend to 1.5
times the interquartile range. Points plotted beyond the whiskers are
considered outliers.

When using categorical variables with linear models, we use boolean
dummy variables to indicate class membership. In the case of
\texttt{edema}, we would probably create two logical variables for
\texttt{edema\ =\ 0.5} (complex Edema presence indicator) and
\texttt{edema\ =\ 1.0} (Edema with diuretics) contrasted with the
\texttt{edema\ =\ 0} variable (no Edema). Random Forest can use factor
variables directly, separating the populations into homogeneous groups
of \texttt{edema} at nodes that split on that variable.
\autoref{fig:variable-plotCat} indicates similar survival response
distribution between 1 and 3 year when \texttt{edema\ =\ 1.0}. The
distribution of predicted survival does seem to spread out more than for
the other values, again indicating a possible non-proportional hazards
response.

\subsection{Partial dependence}\label{partial-dependence}

\emph{Partial dependence} plots are a risk adjusted alternative to
variable dependence. Partial plots are generated by integrating out the
effects of variables beside the covariate of interest. The figures are
constructed by selecting points evenly spaced along the distribution of
the variable of interest. For each of these points (\(X = x\)), we
calculate the average RF prediction over all remaining covariates in the
training set by

\[
\tilde{f}(x) = \frac{1}{n} \sum_{i = 1}^n \hat{f}(x, x_{i, o}),
\label{E:partial}
\]

where \(\hat{f}\) is the predicted response from the random forest and
\(x_{i, o}\) is the value for all other covariates other than \(X = x\)
for observation \(i\) \citep{Friedman:2000}.

Generating partial dependence data is effectively averaging the response
for a series of nomograms constructed for each observation by varying
the variable of interest. The operation is computationally intensive,
especially when there are a large number of observations. The default
parameters for the \texttt{plot.variable} function generate partial
dependence estimates at \texttt{pts\ =\ 25} points along the variable of
interest. For each point of interest, the \texttt{plot.variable}
function averages the \texttt{n} response predictions. This process is
repeated for each of the variables of interest.

For time to event data, we also have to deal with the additional time
dimension, as with variable dependence. The following code block uses
the \texttt{mclapply} function from the \pkg{parallel} package to run
the \texttt{plot.variable} function for three time points
(\texttt{time}=1, 3 and 5 years) in parallel. For RSF models, we
calculate a risk adjusted survival estimates
(\texttt{surv.type="surv"}), suppressing the internal base graphs
(\texttt{show.plots\ =\ FALSE}) and store the point estimates in the
\texttt{partial\_pbc} \texttt{list}.

\begin{Schunk}
\begin{Sinput}
R> xvar <- c(xvar, xvar.cat)
R>
R> time_index <- c(which(rfsrc_pbc$time.interest > 1)[1]-1,
R+                 which(rfsrc_pbc$time.interest > 3)[1]-1,
R+                 which(rfsrc_pbc$time.interest > 5)[1]-1)
R> partial_pbc <- mclapply(rfsrc_pbc$time.interest[time_index],
R+                         function(tm){
R+                           plot.variable(rfsrc_pbc, surv.type = "surv",
R+                                         time = tm, xvar.names = xvar,
R+                                         partial = TRUE ,
R+                                         show.plots = FALSE)
R+                         })
\end{Sinput}
\end{Schunk}

Because partial dependence data is collapsed onto the risk adjusted
response, we can show multiple time curves on a single panel. The
following code block converts the \texttt{plot.variable} output into a
list of \texttt{gg\_partial} objects, and then combines these data
objects, with descriptive labels, along each variable of interest using
the \texttt{combine.gg\_partial} function.

\begin{Schunk}
\begin{Sinput}
R> gg_dta <- mclapply(partial_pbc, gg_partial)
R> pbc_ggpart <- combine.gg_partial(gg_dta[[1]], gg_dta[[2]],
R+                                  lbls = c("1 Year", "3 Years"))
\end{Sinput}
\end{Schunk}

We then segregate the continuous and categorical variables, and generate
a panel plot of all continuous variables in the \texttt{gg\_partial}
plot of \autoref{fig:pbc-partial-panel}. The panels are ordered by
minimal depth ranking. Since all variables are plotted on the same
Y-axis scale, those that are strongly related to survival make other
variables look flatter. The figures also confirm the strong non-linear
contribution of these variables. Non-proportional hazard response is
also evident in at least the \texttt{bili} and \texttt{copper} variables
by noting the divergence of curves as time progresses.

\begin{Schunk}
\begin{figure}[!htb]

{\centering \includegraphics{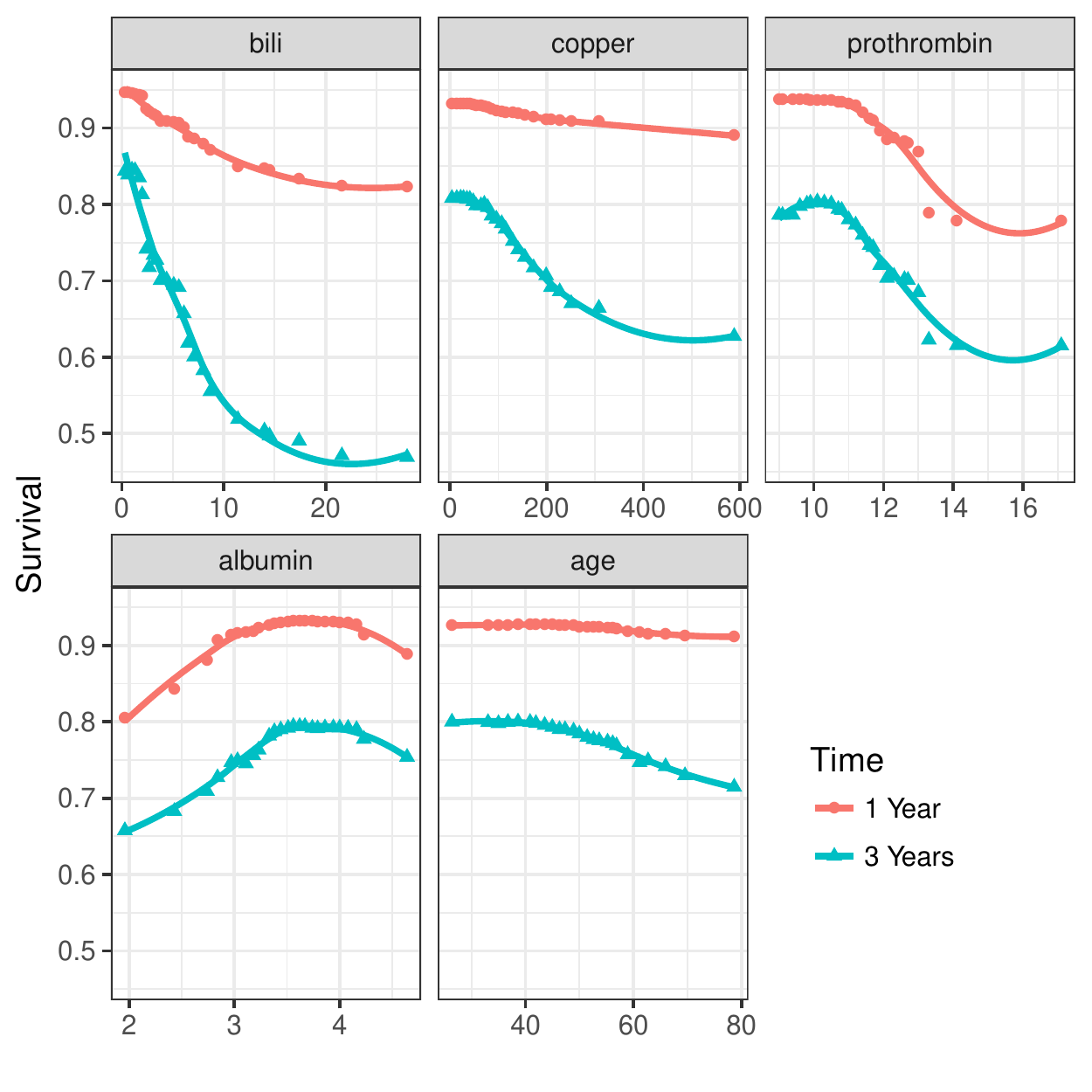}

}

\caption[Partial dependence of predicted survival at 1 year (red circle) and 3 years (blue triangle) as a function continuous variables of interest]{Partial dependence of predicted survival at 1 year (red circle) and 3 years (blue triangle) as a function continuous variables of interest. Symbols are partial dependence point estimates with loess smooth line to indicate trends.}\label{fig:pbc-partial-panel}
\end{figure}
\end{Schunk}

Categorical partial dependence is displayed as boxplots, similar to
categorical variable dependence. Risk adjustment greatly reduces the
spread of the response as expected, and may also move the mean response
compared to the unadjusted results. The categorical \texttt{gg\_partial}
plot of \autoref{fig:pbc-partial-edema} indicates that, adjusting for
other variables, survival decreases with rising \texttt{edema} values.
We also note that the risk adjusted distribution does spread out as we
move further out in time.

\begin{Schunk}
\begin{Sinput}
R> ggplot(pbc_ggpart[["edema"]], aes(y=yhat, x=edema, col=group))+
R+   geom_boxplot(notch = TRUE,
R+                outlier.shape = NA) + # panel=TRUE,
R+   labs(x = "Edema", y = "Survival (%)", color="Time", shape="Time") +
R+   theme(legend.position = c(0.1, 0.2))
\end{Sinput}
\begin{figure}[!htb]

{\centering \includegraphics{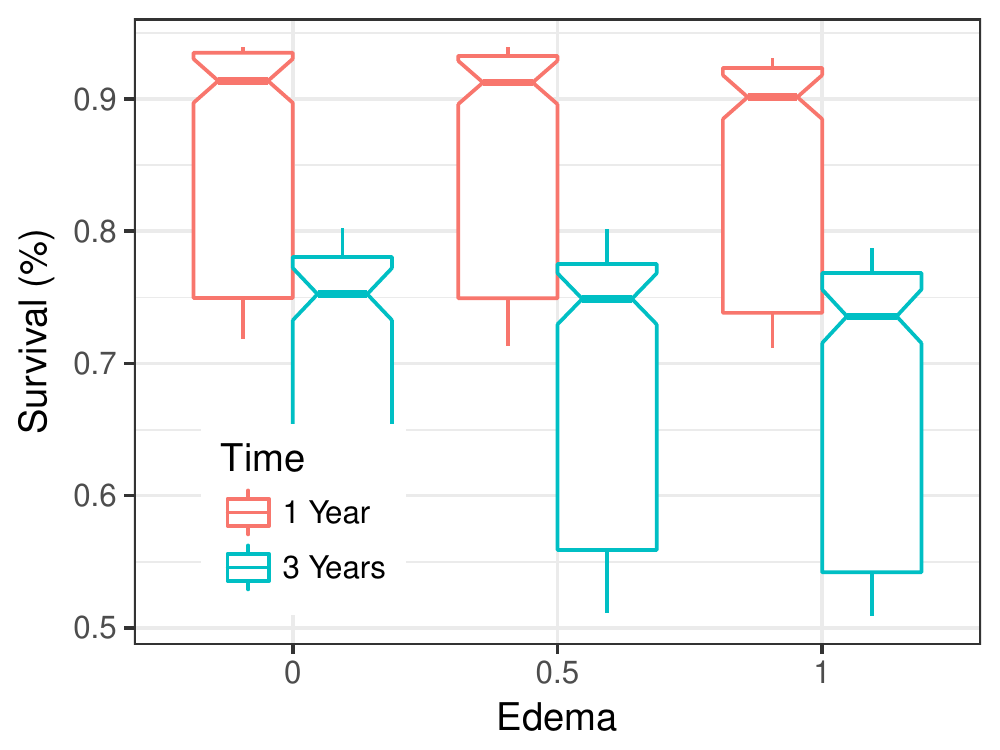}

}

\caption[Partial dependence plot of predicted survival at 1 year (red) and 3 years (blue) as a function of edema groups (categorical variable)]{Partial dependence plot of predicted survival at 1 year (red) and 3 years (blue) as a function of edema groups (categorical variable). Boxplots indicate distribution within each group.}\label{fig:pbc-partial-edema}
\end{figure}
\end{Schunk}

Partial dependence is an extrapolation operation. By averaging over a
series of nomograms, the algorithm constructs observations for all
values of the variable of interest, regardless of the relation with
other variables. In contrast, variable dependence only uses observations
from within the training set. A simple example would be for a model
including BMI, weight and height. When examining partial dependence of
BMI, the algorithm only manipulates BMI values, height or weight values.
The averaging operation is then confounded in two directions. First,
dependence on height and weight is shared with BMI, making it difficult
to see the true response dependence. Second, partial dependence is
calculated over nomograms that can not physically occur. For simple
variable combinations, like BMI, it is not difficult to recognize this
and modify the independent variable list to avoid these issues. However,
care must be taken when interpreting more complex biological variables.

\subsection{Temporal partial
dependence}\label{temporal-partial-dependence}

In the previous section, we calculated risk adjusted (partial)
dependence at two time points (1 and 3 years). The selection of these
points can be driven by biological times of interest (i.e., 1 year and 5
year survival in cancer studies) or by investigating time points of
interest from a \texttt{gg\_rfsrc} prediction plot. We typically
restrict generating \texttt{gg\_partial} plots to the variables of
interest at two or three time points of interest due to computational
constraints.

It is instructive to see a more detailed map of the risk adjusted
response to get a feel for interpreting partial and variable dependence
plots. In \autoref{fig:pbc-partial-panel}, we can visualize the two
curves as extending into the plane of the page along a time axis.
Filling in more partial dependence curves, it is possible to create a
partial dependence surface.

For this exercise, we will generate a series of 50 \texttt{gg\_partial}
plot curves for the \texttt{bili} variable. To fill the surface in, we
also increased the number of points along the distribution of
\texttt{bili} to \texttt{npts=50} to create a grid of \(50 \times 50\)
risk adjusted estimates of survival along time in one dimension and the
\texttt{bili} variable in the second.

\begin{Schunk}
\begin{figure}[!htb]

{\centering \includegraphics{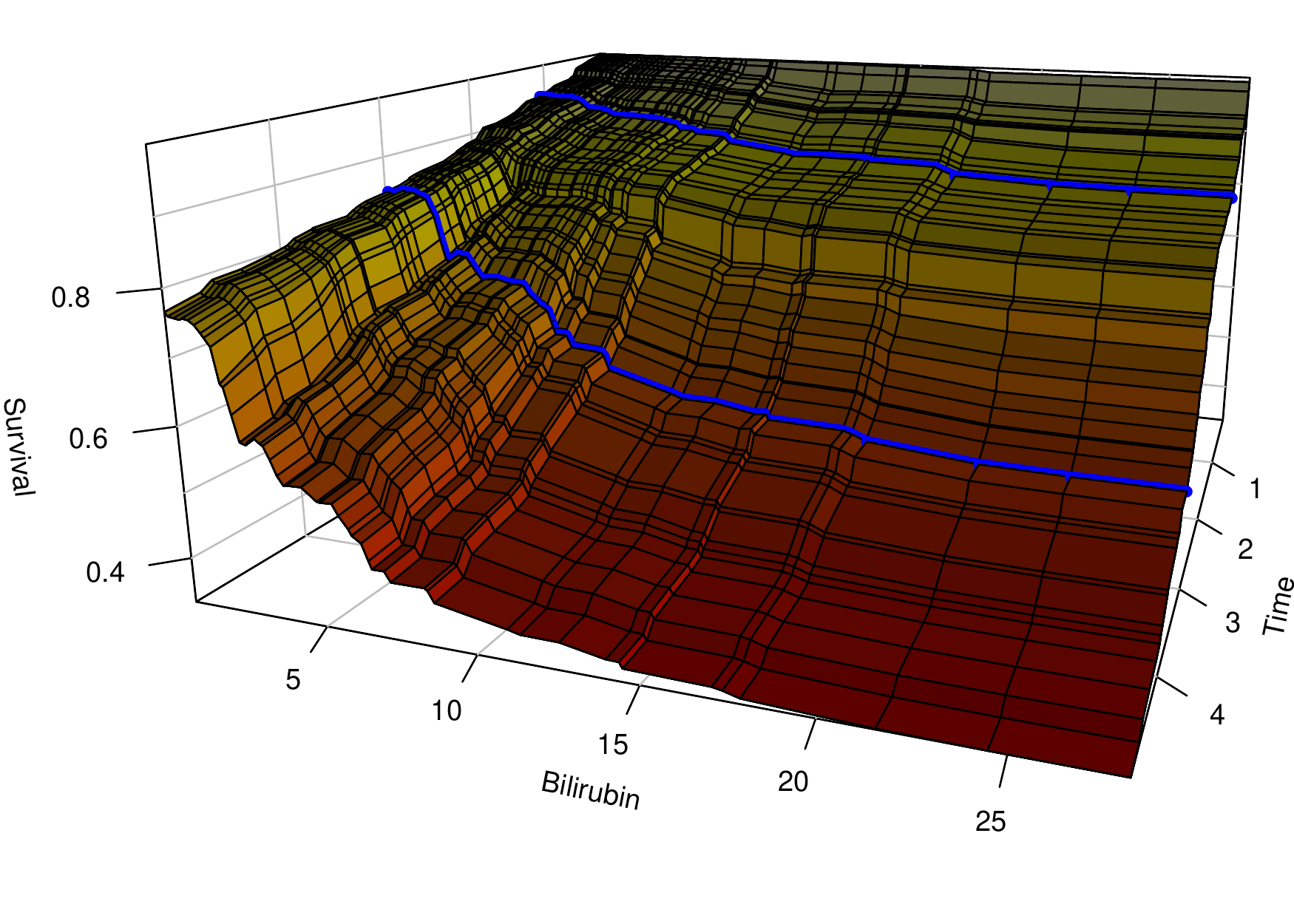}

}

\caption{Partial dependence surface. Partial dependence of predicted survival (0 to 5 years) as a function of bili. Blue lines indicate partial dependence at 1 and 3 years, as in bili panel of \autoref{fig:pbc-partial-panel}.}\label{fig:timeSurface3d}
\end{figure}
\end{Schunk}

The \texttt{gg\_partial} surface of \autoref{fig:timeSurface3d} was
constructed using the \texttt{surf3D} function from the \pkg{plot3D}
package
\citep[\url{http://CRAN.R-project.org/package=plot3D}]{plot3D:2014}.

The figure shows partial dependence of survival (Z-axis) as a function
of \texttt{bili} over a five year follow up time period. Lines
perpendicular to the Bilirubin axis are distributed along the
\texttt{bili} variable. Lines parallel to the Bilirubin axis are taken
at 50 training set event times, the first event after \(t=0\) at the
back to last event before \(t=5\) years at the front. The distribution
of the time lines is also evenly selected using the same procedure as
selecting points for partial dependence curves.

The 2500 estimated partial dependence points are joined together with a
simple straight line interpolation to create the surface, colored
according to the survival estimates (yellow close to 1, red for lower
values) to aid the visualization of 3 dimensions on a 2 dimensional
page. The blue lines in \autoref{fig:timeSurface3d} correspond to the 1
and 3 year partial dependence, as shown in the \texttt{bili} panel of
\autoref{fig:pbc-partial-panel}.

Viewed as a surface, we see how the partial dependence changes with
time. For low values of \texttt{bili}, survival decreases at a constant
rate. For higher values, the rate seems constant until somewhere near 2
years, where it increases rapidly before slowing again as we approach
the 5 year point.

\section{Variable interactions}\label{variable-interactions}

We could stop with the results that our RF analysis has found these six
variables to be important in predicting survival. Where the survival
response is decreasing with increasing \texttt{bili}, \texttt{copper},
\texttt{prothrombin}, \texttt{age} and \texttt{edema} and increasing
with increasing \texttt{albumin}. These results agree with the sign of
the \cite{fleming:1991} model coefficients shown in \autoref{T:FHmodel}.
The \texttt{gg\_partial} plot in \autoref{fig:pbc-partial-panel}
supports the \texttt{log} transform of \texttt{bili}, \texttt{albumin}
and \texttt{prothrombin} and suggest a similar transform for including
the \texttt{copper} variable in a proportional hazards model. The
\texttt{age} variable does seem to have a more linear response than the
other continuous variables, and using dummy variables for \texttt{edema}
would preclude the need for a transformation.

Using minimal depth, it is also possible to calculate measures of
pairwise interactions among variables. Recall that minimal depth measure
is defined by averaging the tree depth of variable \(i\) relative to the
root node. To detect interactions, this calculation can be modified to
measure the minimal depth of a variable \(j\) with respect to the
maximal subtree for variable \(i\) \citep{Ishwaran:2010,Ishwaran:2011}.

The \texttt{randomForestSRC::find.interaction} function traverses the
forest, calculating all pairwise minimal depth interactions, and returns
a \(p \times p\) matrix of interaction measures. The diagonal terms are
normalized to the root node, and off diagonal terms are normalized
measures of pairwise variable interaction.

\begin{Schunk}
\begin{Sinput}
R> ggint <- gg_interaction(rfsrc_pbc)
\end{Sinput}
\begin{Soutput}

                              Method: maxsubtree
                    No. of variables: 17
  Variables sorted by minimal depth?: TRUE

            bili albumin copper prothrombin chol edema platelet sgot  age
bili        0.11    0.26   0.29        0.30 0.29  0.50     0.29 0.31 0.28
albumin     0.33    0.17   0.37        0.39 0.38  0.66     0.37 0.39 0.37
copper      0.36    0.38   0.19        0.40 0.41  0.66     0.41 0.40 0.39
prothrombin 0.40    0.41   0.43        0.20 0.45  0.69     0.46 0.45 0.44
chol        0.42    0.42   0.46        0.48 0.22  0.76     0.47 0.46 0.43
edema       0.45    0.47   0.50        0.51 0.50  0.22     0.49 0.51 0.48
platelet    0.49    0.50   0.53        0.54 0.52  0.80     0.22 0.53 0.51
sgot        0.49    0.47   0.50        0.53 0.52  0.80     0.52 0.25 0.49
age         0.41    0.44   0.47        0.50 0.48  0.79     0.48 0.46 0.25
alk         0.55    0.55   0.57        0.60 0.57  0.83     0.59 0.59 0.55
trig        0.56    0.55   0.57        0.59 0.57  0.84     0.58 0.57 0.55
ascites     0.53    0.54   0.59        0.58 0.58  0.76     0.58 0.60 0.57
stage       0.59    0.60   0.61        0.64 0.61  0.83     0.63 0.62 0.60
hepatom     0.67    0.68   0.70        0.71 0.69  0.86     0.72 0.70 0.69
spiders     0.81    0.82   0.84        0.84 0.84  0.94     0.83 0.82 0.81
treatment   0.87    0.87   0.88        0.87 0.88  0.96     0.89 0.88 0.87
sex         0.84    0.84   0.86        0.86 0.85  0.94     0.86 0.86 0.84
             alk trig ascites stage hepatom spiders treatment  sex
bili        0.31 0.35    0.65  0.45    0.60    0.62      0.61 0.66
albumin     0.39 0.44    0.78  0.55    0.67    0.70      0.66 0.72
copper      0.42 0.45    0.77  0.57    0.69    0.71      0.69 0.79
prothrombin 0.45 0.48    0.81  0.61    0.72    0.76      0.72 0.79
chol        0.46 0.50    0.86  0.63    0.77    0.76      0.72 0.81
edema       0.50 0.55    0.80  0.64    0.73    0.75      0.75 0.81
platelet    0.53 0.55    0.90  0.70    0.81    0.81      0.77 0.86
sgot        0.51 0.53    0.90  0.67    0.79    0.79      0.76 0.84
age         0.46 0.48    0.88  0.62    0.77    0.76      0.72 0.80
alk         0.27 0.61    0.93  0.74    0.85    0.83      0.81 0.88
trig        0.56 0.30    0.92  0.74    0.85    0.84      0.80 0.90
ascites     0.59 0.62    0.34  0.69    0.77    0.80      0.80 0.83
stage       0.62 0.64    0.91  0.35    0.84    0.83      0.81 0.87
hepatom     0.71 0.71    0.91  0.80    0.43    0.88      0.85 0.89
spiders     0.83 0.84    0.96  0.90    0.92    0.53      0.90 0.95
treatment   0.87 0.88    0.99  0.94    0.96    0.95      0.55 0.97
sex         0.86 0.86    0.98  0.91    0.94    0.95      0.92 0.60
\end{Soutput}
\end{Schunk}

The \texttt{gg\_interaction} function wraps the
\texttt{find.interaction} matrix for use with the \pkg{ggRandomForests}
plot and print functions. The \texttt{xvar} argument is used to restrict
the variables of interest and the \texttt{panel\ =\ TRUE} argument
displays the results in a single figure.

\begin{Schunk}
\begin{Sinput}
R> plot(ggint, xvar = xvar)
\end{Sinput}
\begin{figure}[!htb]

{\centering \includegraphics{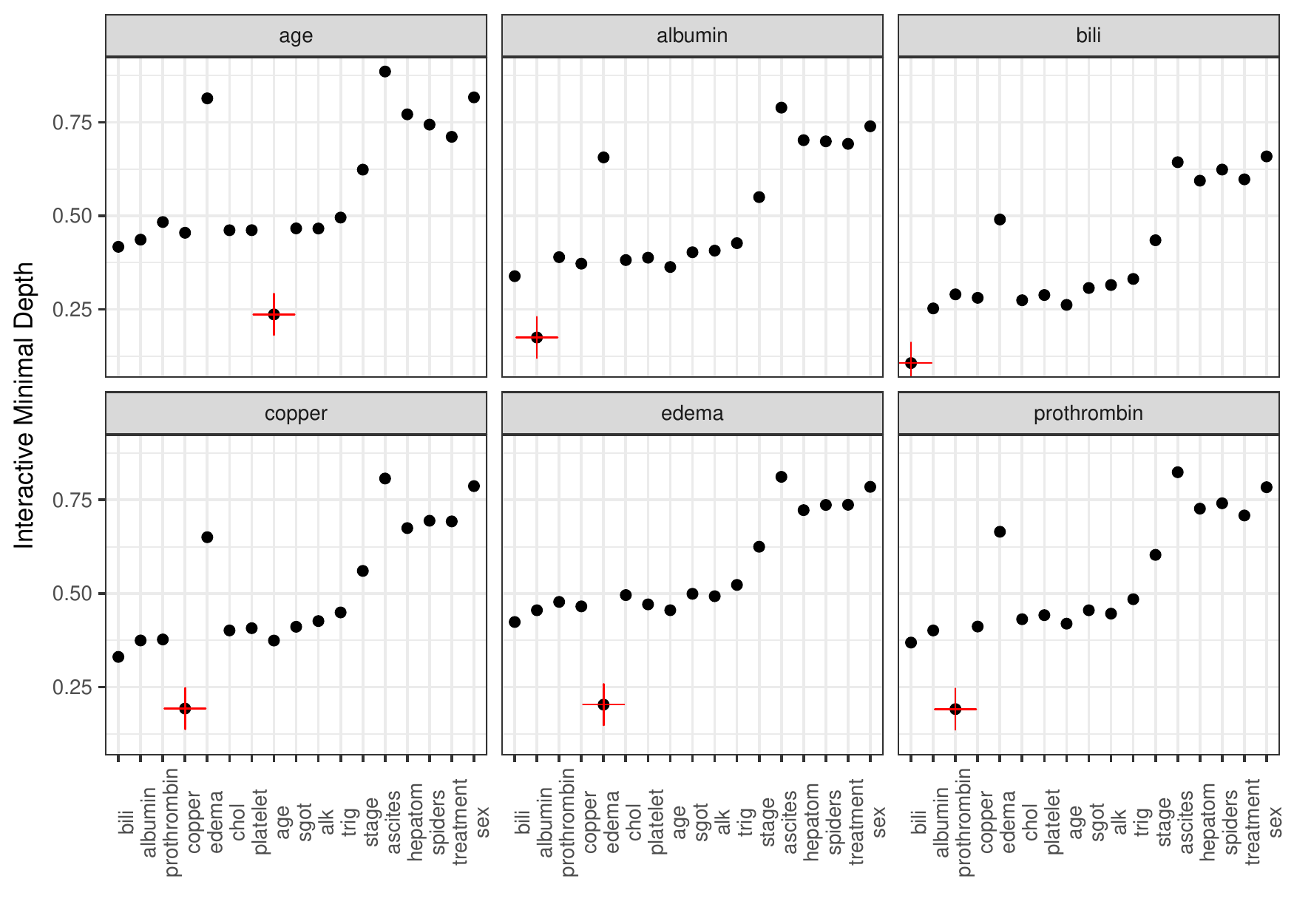}

}

\caption[Minimal depth variable interaction plot for six variables of interest]{Minimal depth variable interaction plot for six variables of interest. Higher values indicate lower interactivity with target variable marked in red.}\label{fig:interactionPanel}
\end{figure}
\end{Schunk}

The \texttt{gg\_interaction} plots in \autoref{fig:interactionPanel}
show interactions for the target variable (shown with the red cross)
with interaction scores for all remaining variables. We expect the
covariate with lowest minimal depth (\texttt{bili}) to be associated
with almost all other variables, as it typically splits close to the
root node, so viewed alone it may not be as informative as looking at a
collection of interactive depth plots. Scanning across the panels, we
see each successive target depth increasing, as expected. We also see
the interactive variables increasing with increasing target depth.

\section{Conditional dependence
plots}\label{conditional-dependence-plots}

Conditioning plots (coplots) \citep{chambers:1992,cleveland:1993} are a
powerful visualization tool to efficiently study how a response depends
on two or more variables \citep{cleveland:1993}. The method allows us to
view data by grouping observations on some conditional membership. The
simplest example involves a categorical variable, where we plot our data
conditional on class membership, for instance on groups of the
\texttt{edema} variable. We can view a coplot as a stratified variable
dependence plot, indicating trends in the RF prediction results within
panels of group membership.

Interactions with categorical data can be generated directly from
variable dependence plots. Recall the variable dependence for bilirubin
shown in \autoref{fig:variable-plotbili}. We recreated the
\texttt{gg\_variable} plot in \autoref{fig:var_dep}, modified by adding
a linear smooth as we intend on segregating the data along conditional
class membership.

\begin{Schunk}
\begin{Sinput}
R> # Get variable dependence at 1 year
R> ggvar <- gg_variable(rfsrc_pbc, time = 1)
R>
R> # For labeling coplot membership
R> ggvar$edema <- paste("edema = ", ggvar$edema, sep = "")
R>
R> # Plot with linear smooth (method argument)
R> var_dep <- plot(ggvar, xvar = "bili",
R+                 alpha = 0.5) +
R+ #  geom_smooth(method = "glm",se = FALSE) +
R+   labs(y = "Survival",
R+        x = st.labs["bili"]) +
R+   theme(legend.position = "none") +
R+   scale_color_manual(values = strCol, labels = event.labels) +
R+   scale_shape_manual(values = event.marks, labels = event.labels) +
R+   coord_cartesian(y = c(-.01,1.01))
R>
R> var_dep
\end{Sinput}
\begin{figure}[!htb]

{\centering \includegraphics{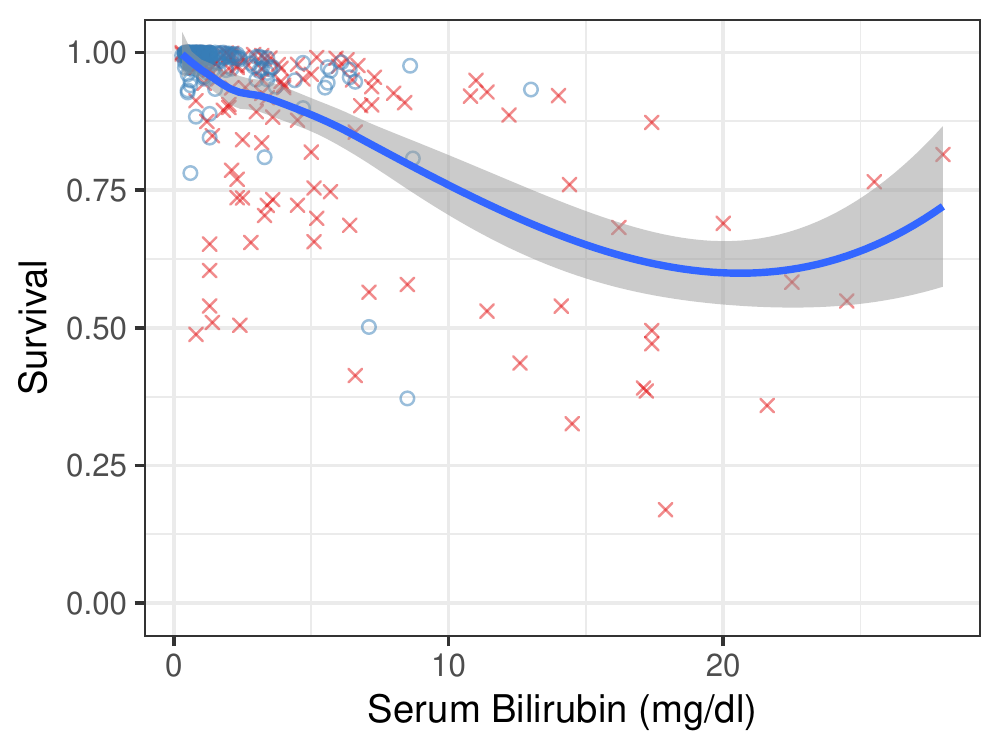}

}

\caption{Variable dependence of survival at 1 year against bili variable. Reproduction of top panel of \autoref{fig:variable-plotbili} with a linear smooth to indicate trend.}\label{fig:var_dep}
\end{figure}
\end{Schunk}

We can view the conditional dependence of survival against bilirubin,
conditional on \texttt{edema} group membership (categorical variable) in
\autoref{fig:coplot_bilirubin} by reusing the saved \texttt{ggplot}
object (\texttt{var\_dep}) and adding a call to the \texttt{facet\_grid}
function.

\begin{Schunk}
\begin{Sinput}
R> var_dep + facet_grid(~edema)
\end{Sinput}
\begin{figure}[!htb]

{\centering \includegraphics{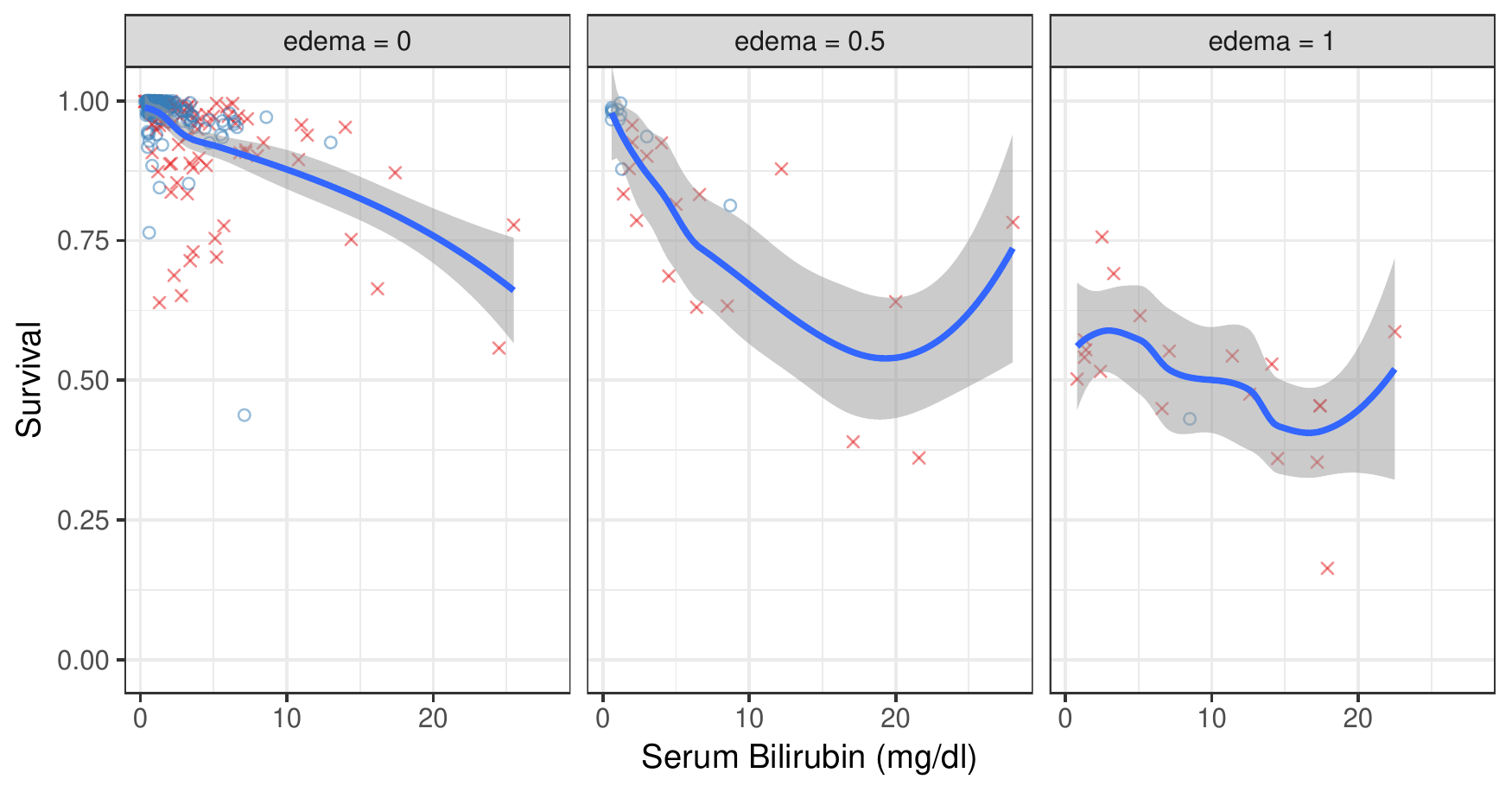}

}

\caption[Variable dependence coplot of survival at 1 year against bili, conditional on edema group membership]{Variable dependence coplot of survival at 1 year against bili, conditional on edema group membership. Linear smooth indicates trend of variable dependence.}\label{fig:coplot_bilirubin}
\end{figure}
\end{Schunk}

Comparing \autoref{fig:var_dep} with conditional panels of
\autoref{fig:coplot_bilirubin}, we see the overall response is similar
to the \texttt{edema=0} response. The survival for \texttt{edema=0.5} is
slightly lower, though the slope of the smooth indicates a similar
relation to \texttt{bili}. The \texttt{edema=1} panel shows that the
survival for this (smaller) group of patients is worse, but still
follows the trend of decreasing with increasing \texttt{bili}.

Conditional membership within a continuous variable requires
stratification at some level. We can sometimes make these stratification
along some feature of the variable, for instance a variable with integer
values, or 5 or 10 year age group cohorts. However with our variables of
interest, there are no logical stratification indications. Therefore we
arbitrarily stratify our variables into 6 groups of roughly equal
population size using the \texttt{quantile\_cuts} function. We pass the
break points located by \texttt{quantile\_cuts} to the \texttt{cut}
function to create grouping intervals, which we can then add to the
\texttt{gg\_variable} object before plotting with the
\texttt{plot.gg\_variable} function. This time we use the
\texttt{facet\_wrap} function to generate the panels grouping interval,
which automatically sorts the six panels into two rows of three panels
each.

\begin{Schunk}
\begin{figure}[!htb]

{\centering \includegraphics{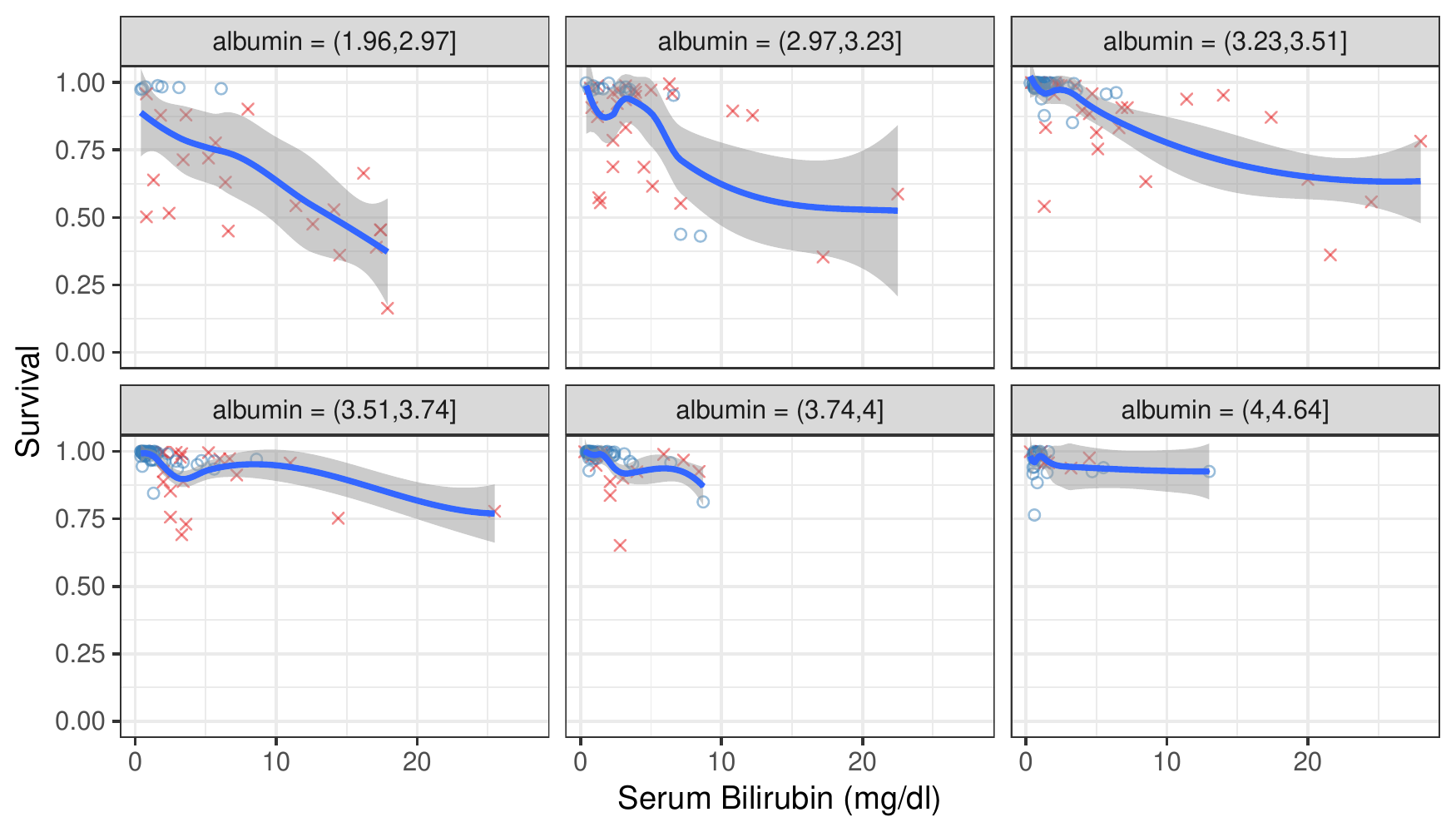}

}

\caption[Variable dependence coplot of survival at 1 year against bili, conditional on albumin interval group membership]{Variable dependence coplot of survival at 1 year against bili, conditional on albumin interval group membership.}\label{fig:albumin-coplot}
\end{figure}
\end{Schunk}

The \texttt{gg\_variable} coplot of \autoref{fig:albumin-coplot}
indicates that the effect of \texttt{bili} decreases conditional on
membership within increasing \texttt{albumin} groups. To get a better
feel for how the response depends on both these variables together, it
is instructive to look at the compliment coplot of \texttt{albumin}
conditional on membership in \texttt{bili} groups. We repeat the
previous coplot process, predicted survival as a function of the
\texttt{albumin} variable, conditional on membership within 6 groups
\texttt{bili} intervals. As the code to create the coplot of
\autoref{fig:bili-coplot} is nearly identical to the code for creating
\autoref{fig:albumin-coplot}.

\begin{Schunk}
\begin{Sinput}
R> # Find intervals with similar number of observations.
R> bili_cts <-quantile_pts(ggvar$bili, groups = 6, intervals = TRUE)
R>
R> # We need to move the minimal value so we include that observation
R> bili_cts[1] <- bili_cts[1] - 1.e-7
R>
R> # Create the conditional groups and add to the gg_variable object
R> bili_grp <- cut(ggvar$bili, breaks = bili_cts)
R> ggvar$bili_grp <- bili_grp
R>
R> # Adjust naming for facets
R> levels(ggvar$bili_grp) <- paste("bilirubin =", levels(bili_grp))
R>
R> # plot.gg_variable
R> plot(ggvar, xvar = "albumin", alpha = 0.5) +
R+ #     method = "glm", se = FALSE) +
R+   labs(y = "Survival", x = st.labs["albumin"]) +
R+   theme(legend.position = "none") +
R+   scale_color_manual(values = strCol, labels = event.labels) +
R+   scale_shape_manual(values = event.marks, labels = event.labels) +
R+   facet_wrap(~bili_grp) +
R+   coord_cartesian(ylim = c(-0.01,1.01))
\end{Sinput}
\begin{figure}[!htb]

{\centering \includegraphics{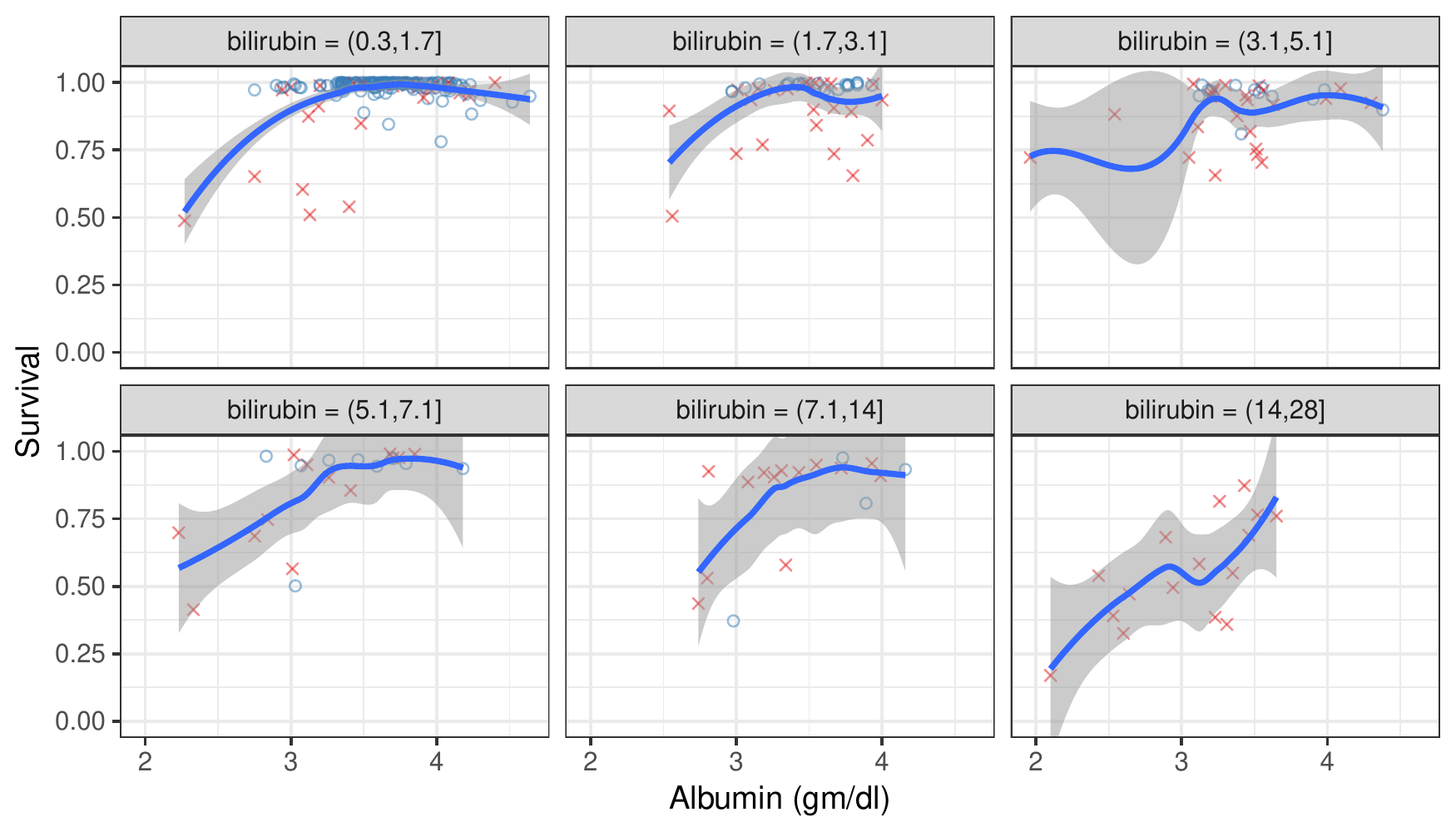}

}

\caption[Variable dependence coplot of survival at 1 year against albumin, conditonal on bili interval group membership]{Variable dependence coplot of survival at 1 year against albumin, conditonal on bili interval group membership.}\label{fig:bili-coplot}
\end{figure}
\end{Schunk}

The \texttt{gg\_variable} coplot of \autoref{fig:bili-coplot} indicates
the probability of survival increases with increasing \texttt{albumin}
and increases within groups of increasing \texttt{bili}.

Typically, conditional plots for continuous variables include
overlapping intervals along the grouped variable \citep{cleveland:1993}.
We chose to use mutually exclusive continuous variable intervals for the
following reasons:

\begin{itemize}
\item
  Simplicity - We can create the coplot figures directly from the
  \texttt{gg\_variable} object by adding a conditional group column
  directly to the object.
\item
  Interpretability - We find it easier to interpret and compare the
  panels if each observation is only in a single panel.
\item
  Clarity - We prefer using more space for the data portion of the
  figures than typically displayed in the \texttt{coplot} function which
  requires the bar plot to present the overlapping segments.
\end{itemize}

It is still possible to augment the \texttt{gg\_variable} to include
overlapping conditional membership with continuous variables by
duplicating rows of the training set data within the
\texttt{rfsrc\$xvar} object, and then setting the conditional group
membership as described. The \texttt{plot.gg\_variable} function recipe
above could be used to generate the panel plot, with panels ordered
according to the factor levels of the grouping variable. We leave this
as an exercise for the reader.

\subsection{Partial dependence
coplots}\label{partial-dependence-coplots}

By characterizing conditional plots as stratified variable dependence
plots, the next logical step would be to generate an analogous
conditional partial dependence plot. The process is similar to variable
dependence coplots, first determine conditional group membership, then
calculate the partial dependence estimates on each subgroup using the
\code{plot.variable} function with a \code{subset} argument for each
grouped interval. The \pkg{ggRandomForests} \code{gg_partial_coplot}
function is a wrapper for generating conditional partial dependence data
objects. Given a random forest (\code{rfsrc}) object and a \code{groups}
vector for conditioning the training data set observations,
\code{gg_partial_coplot} calls the \code{plot.variable} function the
training set observations conditional on \code{groups} membership. The
function returns a \code{gg_partial_coplot} object, a subclass of the
\texttt{gg\_partial} object, which can be plotted with the
\code{plot.gg_partial} function.

The following code block will generate the data object for creating
partial dependence coplot of 1 year survival as a function of
\texttt{bili} conditional on membership within the 6 groups of
\texttt{albumin} intervals that we examined in the
\autoref{fig:albumin-coplot}.

\begin{Schunk}
\begin{Sinput}
R> partial_coplot_pbc <- gg_partial_coplot(rfsrc_pbc, xvar = "bili",
R+                                         groups = ggvar$albumin_grp,
R+                                         surv_type = "surv",
R+                                         time = rfsrc_pbc$time.interest[time_index[1]],
R+                                         show.plots = FALSE)
R>
R> ggplot(partial_coplot_pbc, aes(x=bili, y=yhat, col=group, shape=group)) +
R+   geom_smooth(se = FALSE) +
R+   labs(x = st.labs["bili"], y = "Survival at 1 year (%)",
R+        color = "albumin", shape = "albumin")
\end{Sinput}
\begin{figure}[!htb]

{\centering \includegraphics{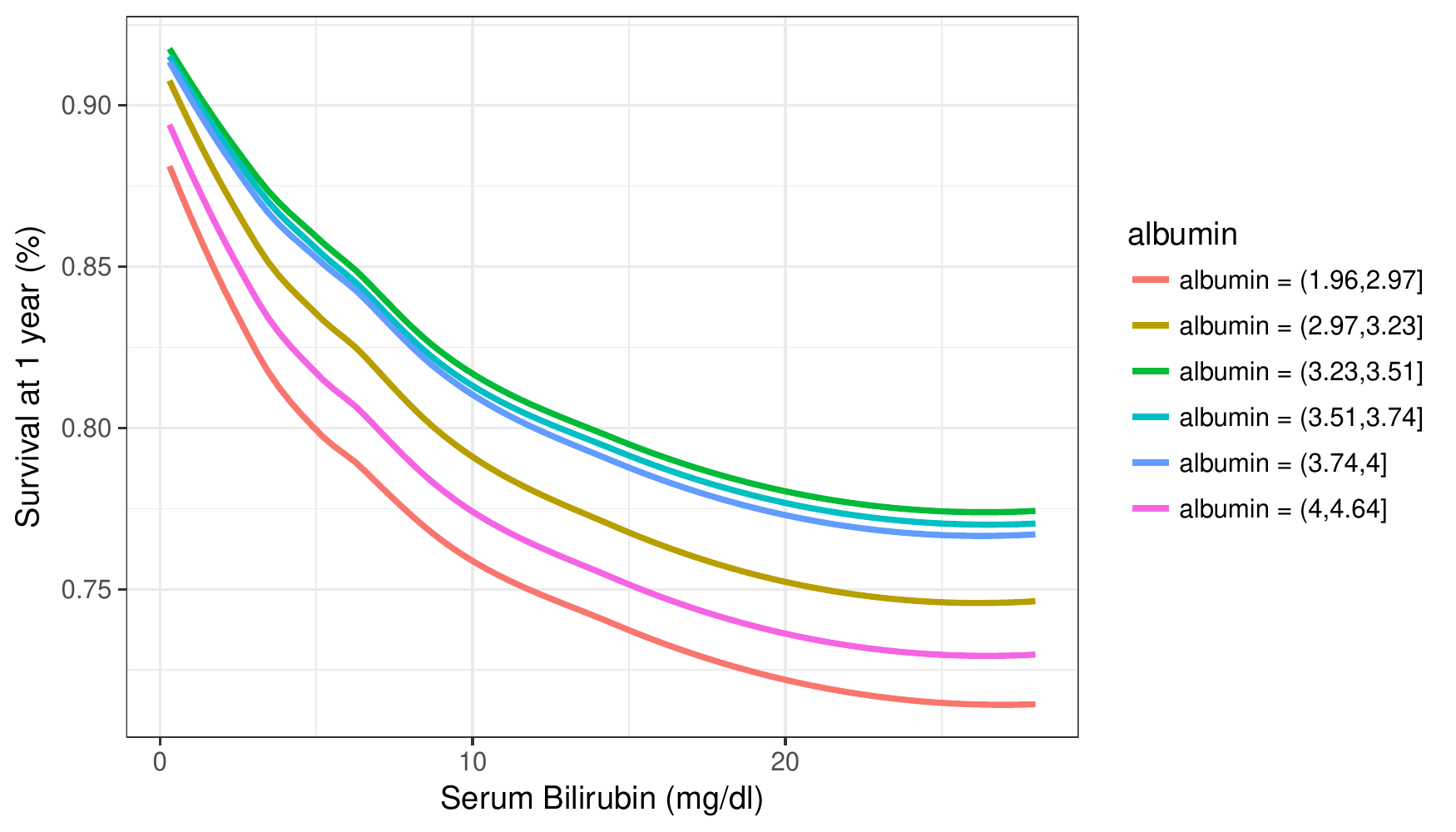}

}

\caption[Partial dependence coplot of survival at 1 year against bili, conditional on albumin interval group membership]{Partial dependence coplot of survival at 1 year against bili, conditional on albumin interval group membership. Points estimates with loess smooth to indicate trend within each group.}\label{fig:bili-albumin}
\end{figure}
\end{Schunk}

The \code{gg_partial_coplot} of \autoref{fig:bili-albumin} shows point
estimates of the risk adjusted survival as a function of \texttt{bili}
conditional on group membership defined by \texttt{albumin} intervals.
The figure is slightly different than the \texttt{gg\_partial} plot of
\autoref{fig:pbc-partial-panel} as each set of partial dependence
estimates is calculated over a subset of the training data. We again
connect the point estimates with a Loess curve.

For completeness, we construct the compliment coplot view of one year
survival as a function of \texttt{albumin} conditional on \texttt{bili}
interval group membership in \autoref{fig:albumin-bili}.

\begin{Schunk}
\begin{figure}[!htb]

{\centering \includegraphics{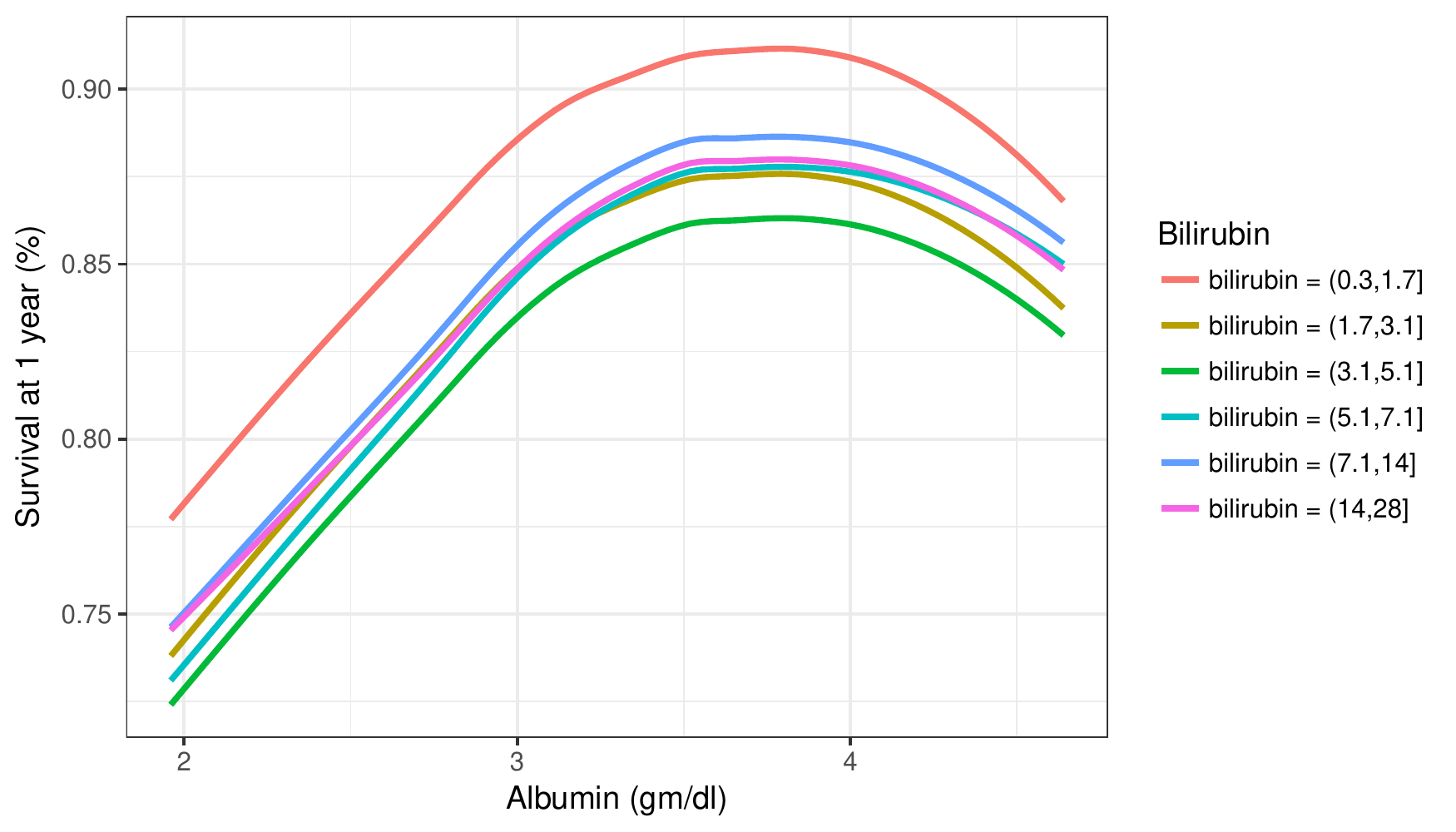}

}

\caption[Partial dependence coplot of survival at 1 year against albumin, conditional on bili interval group membership]{Partial dependence coplot of survival at 1 year against albumin, conditional on bili interval group membership. Points estimates with loess smooth to indicate trend within each group.}\label{fig:albumin-bili}
\end{figure}
\end{Schunk}

\section{Partial plot surfaces}\label{partial-plot-surfaces}

Just as in partial dependence, we can view the partial coplot curves as
slices along a surface that could extend along an axis into the page.
This visualization is made a bit difficult by our choice to select
groups of similar population size, as the curves are not evenly spaced
along the grouping variables. So, similar to the partial dependence
surface we created along time in \autoref{timeSurface}, we can examine
the relation of these two variables using a partial dependence surface.

A difficulty with conditional dependence for this exercise is the
reduction of the sample sizes for calculating a coplot surface. So
instead, we calculate the full partial dependence surface by generating
50 \texttt{albumin} values spaced evenly along the data distribution.
For each value of \texttt{albumin}, we calculate the partial dependence
on \texttt{bili} at \texttt{npts\ =\ 50} points with the
\texttt{plot.variable} function. We generate the surface again using the
\texttt{surf3D} function.

\begin{Schunk}
\begin{figure}[!htb]

{\centering \includegraphics{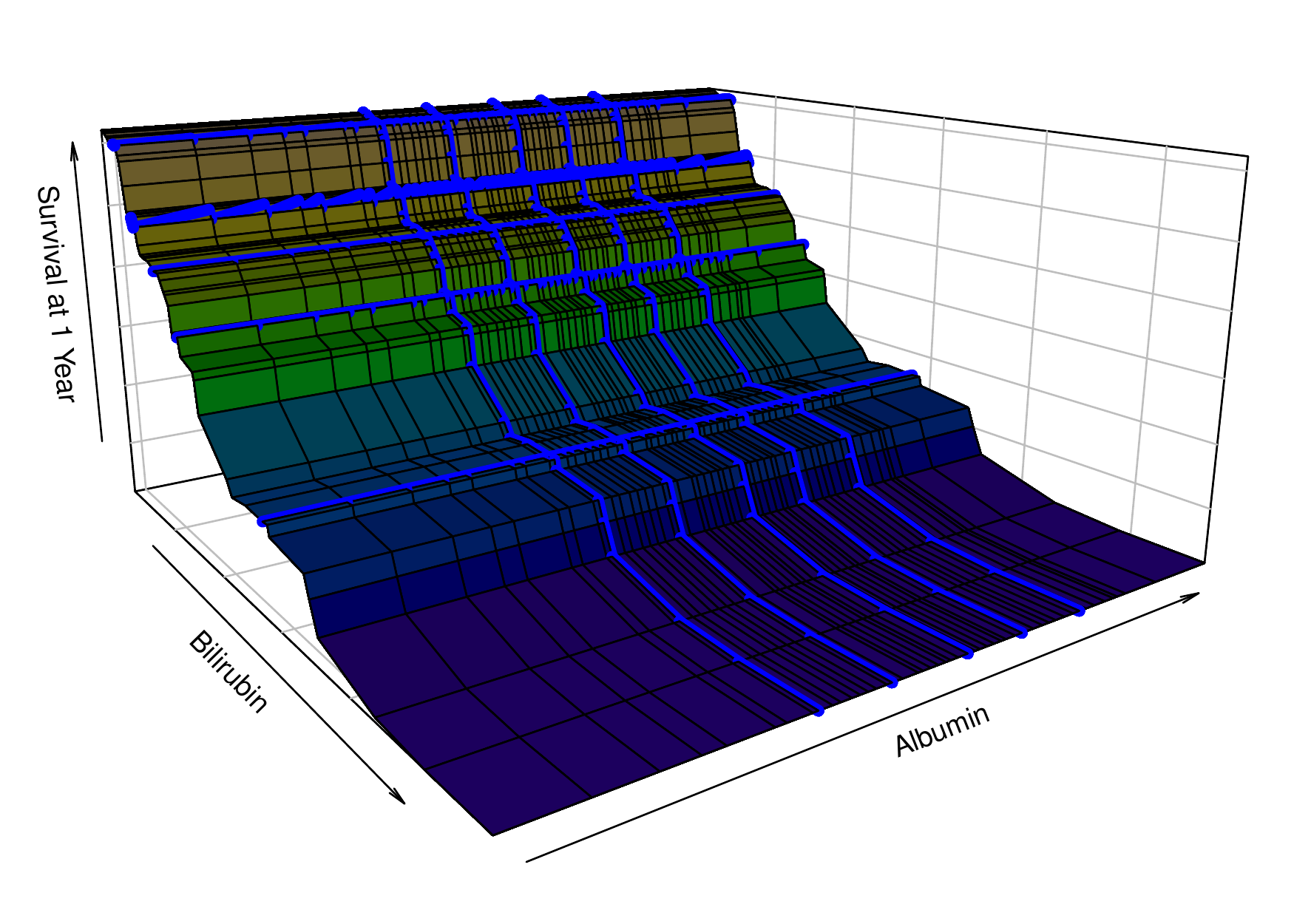}

}

\caption{Partial dependence surface of survival at 1 year as a funtion of bili and albumin. Blue lines indicate partial coplot cut points for albumin (\autoref{fig:bili-albumin}) and bili (\autoref{fig:albumin-bili}).}\label{fig:surface3d}
\end{figure}
\end{Schunk}

The partial dependence surface of \autoref{fig:surface3d} shows partial
dependence of 1 year survival on the Z-axis against values of Bilirubin
and Albumin. We again use linear interpolation between the 2500
estimates, and color the surface by the response. Here blue corresponds
to lower and yellow to higher risk adjusted survival. The blue lines are
placed at the cut points between groups of \texttt{albumin} and
\texttt{bili} used in the partial coplots of Figures
\ref{fig:bili-albumin} and \ref{fig:albumin-bili} respectively.

To construct the partial coplot for groups of \texttt{albumin} in
\autoref{fig:bili-albumin}, we arbitrarily segmented the training set
into 6 groups of equal membership size. The segments between blue lines
parallel to the Bilirubin axis indicate where on the surface these
observations are located. Similarly, the blues lines perpendicular to
the Bilirubin axis segment observations into the 6 groups of
\texttt{bili} intervals. \autoref{fig:surface3d} indicates the arbitrary
grouping for groups of \texttt{bili} in \autoref{fig:albumin-bili}.

The figure indicates that partial dependence of higher \texttt{albumin}
levels are similar, which results in the over plotting seen in
\autoref{fig:bili-albumin}. The distribution is sparser at lower
\texttt{albumin} levels, creating the larger area in lowest
\texttt{albumin} values, where the partial dependence changes the most.

\section{Conclusion}\label{conclusion}

In this vignette, we have demonstrated the use of Random Survival Forest
methods with the
\pkg{ggRandomForests}\textasciitilde{}(\url{http://CRAN.R-project.org/package=ggRandomForests})
package. We have shown how to grow a random forest model and determine
which variables contribute to the forest prediction accuracy using both
VIMP and Minimal Depth measures. We outlined how to investigate variable
associations with the response variable using variable dependence and
the risk adjusted partial dependence plots. We've also explored variable
interactions by using pairwise minimal depth interactions and directly
viewed these interactions using variable dependence coplots and partial
dependence coplots. Along the way, we've demonstrated the use of
additional commands from the \pkg{ggplot2} package
\citep[\url{http://CRAN.R-project.org/package=ggplot2}]{Wickham:2009}
package for modifying and customizing plots from \pkg{ggRandomForests}
functions.

\subsection{Computational details}\label{computational-details}

This document is a package vignette for the \pkg{ggRandomForests}
package for ``Visually Exploring Random Forests''
(\url{http://CRAN.R-project.org/package=ggRandomForests}). The
\pkg{ggRandomForests} package is designed for use with the
\pkg{randomForestSRC} package
\citep[\url{http://CRAN.R-project.org/package=randomForestSRC}]{Ishwaran:RFSRC:2014}
for growing survival, regression and classification random forest models
and uses the \pkg{ggplot2} package
\citep[\url{http://CRAN.R-project.org/package=ggplot2}]{Wickham:2009}
for plotting diagnostic and variable association results.
\pkg{ggRandomForests} is structured to extract data objects from
\pkg{randomForestSRC} objects and provides functions for printing and
plotting these objects.

The vignette is a tutorial for using the \pkg{ggRandomForests} package
with the \pkg{randomForestSRC} package for building and post-processing
random survival forests. In this tutorial, we explore a random forest
for survival model constructed for the primary biliary cirrhosis (PBC)
of the liver data set \citep{fleming:1991}, available in the
\pkg{randomForestSRC} package. We grow a random survival forest and
demonstrate how \pkg{ggRandomForests} can be used when determining how
the survival response depends on predictive variables within the model.
The tutorial demonstrates the design and usage of many of
\pkg{ggRandomForests} functions and features and also how to modify and
customize the resulting \texttt{ggplot} graphic objects along the way.

The vignette is written in \LaTeX using the \pkg{knitr} package
\citep[\url{http://CRAN.R-project.org/package=knitr}]{Xie:2015, Xie:2014,Xie:2013},
which facilitates weaving \proglang{R} \citep{rcore} code, results and
figures into document text.

This vignette is available within the \pkg{ggRandomForests} package on
the Comprehensive R Archive Network (CRAN)
\citep[\url{http://cran.r-project.org}]{rcore}. Once the package has
been installed, the vignette can be viewed directly from within
\proglang{R} with the following command:

\begin{Schunk}
\begin{Sinput}
R> vignette("randomForestSRC-Survival", package = "ggRandomForests")
\end{Sinput}
\end{Schunk}

A development version of the \pkg{ggRandomForests} package is also
available on GitHub (\url{https://github.com}). We invite comments,
feature requests and bug reports for this package at
\url{https://github.com/ehrlinger/ggRandomForests}.

\subsection{Acknowledgement}\label{acknowledgement}

This work was supported in part by the National Institutes of Health
grant R01-HL103552-01A1.

\renewcommand\refname{References}
\bibliography{ggRandomForests}

\end{document}